\documentclass[prd,aps,showpacs,showkeys]{revtex4}

\usepackage{amssymb}
\usepackage{amsmath}
\usepackage{latexsym}

\usepackage[dvips]{graphicx}

\newcommand{\sikib}{\begin{eqnarray}}
\newcommand{\sikie}{\end{eqnarray}}
\newcommand{\sikibnon}{\begin{eqnarray*}}
\newcommand{\sikienon}{\end{eqnarray*}}

\newcommand{\sigmap}{\sigma^{\prime}}
\newcommand{\gin}[1]{g_{in}(\vec{#1})}
\newcommand{\gout}[1]{g_{out}(\vec{#1})}
\newcommand{\gb}[1]{g_g(\vec{#1})}
\newcommand{\gh}[1]{g_h(\vec{#1})}
\newcommand{\jn}{J_{npt}}
\newcommand{\je}{J_{empty}}
\newcommand{\tn}{t_{npt}}
\newcommand{\te}{t_{empty}}

\begin{document}

\title{Black hole evaporation in a heat bath as a nonequilibrium process and its final fate}

\author{Hiromi Saida}

\email{saida@daido-it.ac.jp}

\affiliation{Department of Physics, Daido Institute of Technology,\\
Takiharu-cho 10-3, Minami-ku, Nagoya 457-8530, Japan}


\begin{abstract}
We consider a black hole in a heat bath, and the whole system which consists of the black hole and the heat bath is isolated from outside environments. When the black hole evaporates, the Hawking radiation causes an energy flow from the black hole to the heat bath. Therefore, since no energy flow arises in an equilibrium state, thermodynamic state of the whole system is not in an equilibrium. That is, in a region around the black hole, the matter field of Hawking radiation and that of heat bath should be in a nonequilibrium state due to the energy flow. Using a simple model which reflects the nonequilibrium nature of energy flow, we find the nonequilibrium effect on a black hole evaporation as follows: if the nonequilibrium region around black hole is not so large, the evaporation time scale of black hole in a heat bath becomes longer than that in an empty space (a situation without heat bath), because of the incoming energy flow from the heat bath to the black hole. However, if the nonequilibrium region around black hole is sufficiently large, the evaporation time scale in a heat bath becomes shorter than that in an empty space, because a nonequilibrium effect of temperature difference between the black hole and the heat bath appears as a strong energy extraction from the black hole by the heat bath. Further a specific nonequilibrium phenomenon is found: a quasi-equilibrium evaporation stage under the nonequilibrium effect proceeds abruptly to a quantum evaporation stage at semi-classical level (at black hole radius $R_g >$ Planck length) within a very short time scale with a strong burst of energy. (Contrary, when the nonequilibrium effect is not taken into account, a quasi-equilibrium stage proceeds smoothly to a quantum stage at $R_g <$ Planck length without so strong energy burst.) That is, the nonequilibrium effect of energy flow tends to make a black hole evaporation process more dynamical and to accelerate that process. Finally on the final fate of black hole evaporation, we find that, in order to make total entropy of the whole system increase along an evaporation process, a remnant should remain after the evaporation of black hole without respect to the size of nonequilibrium region around black hole. This implies that the information loss problem may disappear due to the nonequilibrium effect of energy flow.
\end{abstract}


\keywords{Black hole evaporation, Nonequilibrium thermodynamics, Black hole phase transition, Information loss problem \\[1mm]
Published in {\it Classical and Quantum Gravity} {\bf 24} (2007) 691 -- 722}

\maketitle

\section{Introduction}
\label{sec-intro}

Consider that a black hole is put in a heat bath, and that the whole system, which consists of the black hole and the heat bath, is isolated from outside environments. It has already been known that, by considering only the equilibrium states of the whole system, the equilibrium state is unstable for a sufficiently small black hole, and stable for a sufficiently large black hole \cite{ref-trans.1} \cite{ref-trans.2}. If the instability occurs for a small black hole and the system starts to evolve toward the other stable state, there are two possibilities of its evolution. The first is that, due to the statistical (and/or quantum) fluctuation, the temperature of heat bath exceeds that of black hole. Then the black hole swallows a part of heat bath and settles down to a stable equilibrium state of a larger black hole in a heat bath. The second possible evolution is that, due to the statistical (and/or quantum) fluctuation, the temperature of heat bath becomes lower than that of black hole. Then the black hole emits its mass energy by the Hawking radiation and settles down to some other stable state. However we do not know the details of end state of the second possibility, since the final fate of black hole evaporation is an unresolved issue at present. A more detailed explanation is given in section \ref{sec-eq}.

When one distinguishes the phase of the system by a criterion whether a black hole can exist stably in an equilibrium with a heat bath or not, the phase transition of the system occurs in varying the black hole radius. This phenomenon is known as the black hole phase transition \cite{ref-trans.1} \cite{ref-trans.2}
\footnote{The phase transition of AdS black holes is known as the Hawking-Page phase transition. However we do not consider a cosmological constant in this paper.}.
This will be important in the study of primordial black holes \cite{ref-pbh}, since the black holes would be formed in a radiation field in the early universe. And, in relation with the possibility of black hole creation in particle accelerators in the context of TeV gravity \cite{ref-tev}, the black hole phase transition will be important as well, since a black hole should be created in a quark-gluon plasma in our brane and in a graviton gas in the higher dimensional spacetime. Therefore, although the black hole phase transition is an old and rather well studied topic, a more study on the black hole phase transition may give some contribution to the present trend in the study of black hole physics.

As mentioned in the first paragraph, if an instability occurs for an equilibrium of a black hole with a heat bath, the black hole may evaporate. If the mass energy of black hole is radiated out completely by the Hawking radiation, then the thermal spectrum of Hawking radiation implies that only a matter field which is in a thermal equilibrium state may be left after the evaporation. This implies that the initial condition of black hole formation (gravitational collapse) is completely smeared out. For example, even if the initial state of collapsing matter is a pure quantum state, the final state after a black hole evaporation must be transformed to a thermal state. This contradicts the unitary invariance of quantum theory. This issue is known as the information loss problem \cite{ref-info}. The study on black hole phase transition, especially on a black hole evaporation after an instability of equilibrium occurs, may give some insight into the information loss problem.

In this paper, we concentrate on a black hole evaporation in a heat bath after an instability of equilibrium occurs. During the evaporation process, there should exist an energy flow from the black hole to the heat bath due to the Hawking radiation. Since no energy flow arises in an equilibrium state, thermodynamic state of the whole system is not in an equilibrium. That is, in a region around a black hole, the matter field of Hawking radiation and that of heat bath should be in a nonequilibrium state due to the energy flow. However it should be noted here that the study on nonequilibrium phenomena is one of the most difficult subjects in physics, and it is impossible at present to treat the nonequilibrium nature of black hole evaporation in a full general relativistic framework. Hence we resort to a simplified model which reflects the nonequilibrium nature of black hole evaporation. Then we aim to find some insight into nonequilibrium effects on a black hole evaporation in a heat bath, and on the final fate of black hole evaporation.

One may think that the generalised second law should be one of interesting issues concerning a black hole evaporation. For the case of a black hole evaporation in an empty space (a situation without heat bath), the generalised second law has already been discussed in reference \cite{ref-gsl} with taking the nonequilibrium nature of black hole evaporation into account. The essence of the generalised second law in nonequilibrium situations has already been revealed in reference \cite{ref-gsl}. Hence the generalised second law is not in the scope of this paper. This paper may be considered as an extension of reference \cite{ref-gsl}, although this paper is written independently of reference \cite{ref-gsl}.

The plan of this paper is as follows. Section \ref{sec-eq} reviews the original discussion of black hole phase transition given in reference \cite{ref-trans.1}, which is based only on equilibrium states of the system. Then section \ref{sec-ne} is devoted to discussions about nonequilibrium effects of energy flow between a black hole and a heat bath, and insights into the nonequilibrium effects are presented. Summary and discussions are in section \ref{sec-sd}.

Throughout this paper, we use the Planck unit, $\hbar = c = G = k_B = 1$. Then the Stefan-Boltzmann constant becomes $\sigma = \pi^2/60$, which is appropriate for a photon gas. When one consider a general non-self-interacting massless matter fields, as indicated in appendix \ref{app-sst}, it is necessary to replace the Stefan-Boltzmann constant as $\sigma \to \sigmap = N\, \pi^2/120 = N\, \sigma/2$, where $N = n_b + ( 7/8 )\, n_f$, and $n_b$ is the number of helicities of massless bosonic fields and $n_f$ is that of massless fermionic fields. ($n_b = 2$ for photons.)

\section{Equilibrium model}
\label{sec-eq}

Before proceeding to the study of nonequilibrium effects of energy flow between a black hole and a heat bath, we briefly review the original discussion of black hole phase transition given in reference \cite{ref-trans.1}. This is based only on equilibrium states of the whole system which consists of a black hole and a heat bath. For simplicity we consider a Schwarzschild black hole in a heat bath.

According to the black hole thermodynamics \cite{ref-bht} \cite{ref-hr}, a Schwarzschild black hole can be modelled as a spherical black body whose equations of states are
\sikib
 E_g = \frac{1}{8 \pi T_g} = \frac{R_g}{2} \quad , \quad
 S_g = \frac{1}{16 \pi T_g^2} = \pi R_g^2 \, ,
\label{eq-eq.eos}
\sikie
where $R_g$, $T_g$, $S_g$ are respectively the radius, temperature and entropy of the body, and $E_g$ is the internal energy of the body which corresponds to the mass of a black hole. It is obvious from these equations that the radius $R_g$ decreases when the body loses its energy $E_g$. The black hole evaporation can be represented by the energy loss of this body. Further note that the heat capacity $C_g$ of this body is negative,
\sikib
 C_g = \frac{dE_g}{dT_g} = - \frac{1}{8 \pi T_g^2} = - 2 \pi R_g^2 < 0 \, .
\label{eq-eq.capacity}
\sikie
The negative heat capacity is a peculiar property of self-gravitating systems \cite{ref-sgs}. That is, the energy $E_g$ includes self-gravitational effects of a black hole on its own thermodynamic state. Further it has already been revealed that \cite{ref-entropy}, using the Euclidean path-integral method for a black hole spacetime and matter fields on it, the entropy of the whole gravitational field of a black hole spacetime is given by the quantity $S_g$ in equations of states \eqref{eq-eq.eos}. That is, the entropy $S_g$ includes not only the entropy of black hole but also the entropy of gravitational field outside the black hole. Hence it should be emphasised here that energetic and entropic properties of a black hole are encoded in the equations of states \eqref{eq-eq.eos}, and we call the body {\it the black hole}.

Put a black hole into a heat bath of heat capacity $C_h^{(eq)} > 0$, and set that the whole system which consists of the black hole and the heat bath is isolated from outside environments and in an equilibrium state (``micro-canonical ensemble''). The temperature of heat bath $T_h$ equals $T_g$. We set $C_h^{(eq)} = constant \,( > 0 )$ for simplicity. The statistical (and/or quantum) fluctuation raises a temperature difference $\delta T = T_g - T_h \neq 0$. Then, there are two possibilities for the evolution of the system after the fluctuation occurs. The evolution is determined by the size of black hole and the signature of $\delta T$.

Consider the case $\left| C_g \right| > C_h^{(eq)}$. If the initial fluctuation is $\delta T > 0$, the energy flows from the black hole to the heat bath due to the second law of ordinary thermodynamics, and relations $dE_g < 0$ and $dE_h > 0$ hold, where $E_h$ is the energy of heat bath. Then $dT_g > 0$ and $dT_h > 0$ hold due to $C_g < 0$ and $C_h > 0$. Further, because the whole system is isolated, $dE_g + dE_h = 0$ holds, and $\left| dE_g \right| = dE_h$ follows. Therefore we find $dT_g = dE_g/C_g < dE_h/C_h^{(eq)} = dT_h \, \Rightarrow \, dT_g < dT_h$. This means that $\delta T \to 0$ after a sufficiently long time. The same discussion holds for the initial fluctuation $\delta T < 0$. That is, the equilibrium state of the whole system is stable for a sufficiently large black hole which satisfies $\left| C_g \right| = 2 \pi R_g^2 > C_h^{(eq)}$.

Consider the case $\left| C_g \right| < C_h^{(eq)}$. If the initial fluctuation is $\delta T > 0$, the energy flows from the black hole to the heat bath, $dE_g < 0$ and $dE_h > 0$. Then $dT_g > 0$ and $dT_h > 0$ hold due to $C_g < 0$ and $C_h > 0$. However, the same calculations as in the previous paragraph result in inequality, $dT_g > dT_h$. This means $\delta T \to \infty$ and $R_g \to 0$ after a sufficiently long time, which corresponds to a black hole evaporation. On the other hand, for the initial fluctuation $\delta T < 0$, we obtain $dT_g < 0$, $dT_h < 0$ and $\left| dT_g \right| > \left| dT_h \right|$. This means that the black hole grows $dR_g > 0$, and finally the heat capacity becomes to satisfy the stable condition $\left| C_g \right| > C_h^{(eq)}$. In summary for the case $\left| C_g \right| < C_h^{(eq)}$, the whole system transfers to a stable equilibrium state with sufficiently large black hole for a fluctuation $\delta T < 0$, or to some other stable state without a black hole (end state of black hole evaporation) for a fluctuation $\delta T >0$. That is, the equilibrium state of the whole system is unstable for a sufficiently small black hole which satisfies $\left| C_g \right| < C_h^{(eq)}$.

Hence from the above, we find that the phase transition of the whole system occurs in varying the black hole radius, where the phase of the system is distinguished by the criterion whether a black hole can exist stably in an equilibrium with a heat bath or not. This phenomenon is known as {\it the black hole phase transition}.

The above discussion is primitive. Although a self-gravitational effect of a black hole on its own thermodynamic state is encoded in equations of states \eqref{eq-eq.eos}, the above discussion ignores gravitational interactions between a heat bath and a black hole. However, using the Euclidean path-integral method for a black hole spacetime and matter fields on it, reference \cite{ref-trans.2} has extended this primitive discussion to include the gravitational interactions, then the same result (stable equilibrium for a large black hole and unstable equilibrium for a small black hole) has been obtained, although reference \cite{ref-trans.2} considers the case that the temperature of heat bath is kept constant (``canonical ensemble''). Here note that thermodynamic properties of an equilibrium state can be derived by statistical mechanics with using either a micro-canonical ensemble or a canonical ensemble. Hence, the difference of ensembles does not become an essential difference between references \cite{ref-trans.1} and \cite{ref-trans.2}. That is, the primitive discussion given in reference \cite{ref-trans.1} retains the essence of black hole phase transition. An intuitive understanding why reference \cite{ref-trans.1} retains the essence is given in appendix \ref{app-essence}. We can find an understanding by references \cite{ref-trans.1} and \cite{ref-trans.2} and by appendix \ref{app-essence}; it is not the gravitational interactions between black hole and its environments but the self-gravitational effects of black hole on its own thermodynamic state (e.g. the negative heat capacity) that causes the black hole phase transition. In next section, we put our basis on reference \cite{ref-trans.1} in taking the nonequilibrium nature of black hole evaporation into account, and consider an isolated case of the whole system.

\section{Nonequilibrium effects}
\label{sec-ne}

\subsection{Simple model of a black hole in a heat bath}
\label{sec-ne.model}

The discussion so far has been based only on equilibrium states of the whole system which consists of a black hole and a heat bath. However as mentioned in section \ref{sec-intro}, there should be a nonequilibrium region around a black hole due to the energy flow which is caused by the energy exchange between the black hole and the heat bath. The energy exchange is never carried by a heat conduction but carried by an exchange of particle fluxes, since a black hole is not composed of matters but is a certain spacetime region \cite{ref-he}. The particle flux from the black hole is the Hawking radiation, and that into the black hole is the accretion of matters emitted by the heat bath. Hereafter, for simplicity and as explained at the end of appendix \ref{app-essence}, we consider general non-self-interacting massless matter fields as the matters which carry the energy exchange between the black hole and the heat bath. For example, photon, graviton, neutrino (if it is massless) and free Klein-Gordon field ($\Box \Phi = 0$) are candidates of such matter fields. Note that these fields possess the generalised Stefan-Boltzmann constant $\sigmap$ as mentioned at the end of section \ref{sec-intro}.

\begin{figure}[t]
 \begin{center}
 \includegraphics[height=30mm]{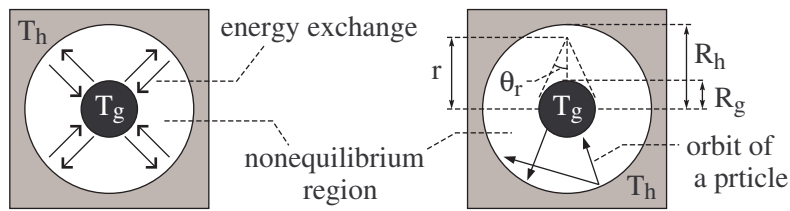}
 \end{center}
\caption{NPT model. Left panel shows the particle flux between a black hole and a heat bath. Right one shows some variables and particles of radiation fields emitted by black hole and heat bath.}
\label{fig-1}
\end{figure}

It is very difficult at present to treat nonequilibrium phenomena around a black hole in a full general relativistic framework. Therefore, in order to pick up the nonequilibrium effect of energy exchange between a black hole and a heat bath, we use a simple model which is obtained by modifying the equilibrium model considered in section \ref{sec-eq}. We remove a shell region around the black hole from the heat bath. Then the space of the removed region between black hole and heat bath is filled up with matter fields emitted by the black hole and the heat bath. Here note that, as explained in appendix \ref{app-essence}, when we use thermodynamic quantities which are evaluated at asymptotically flat region on a black hole spacetime, the thermodynamic property of the whole system which consists of a black hole, a heat bath and matter fields between them may be described by the model which ignores gravitational interactions among a black hole, a heat bath and matter fields. Hence, we consider the model named NPT after the Nonequilibrium black hole Phase Transition.
\begin{description}
\item[NPT model:]
Put a spherical black body of temperature $T_g$ in a heat bath of temperature $T_h (\neq T_g)$, and set that the whole system is isolated from outside environments. Consider the case that the equations of states of the black body are given by equations \eqref{eq-eq.eos}, and that the black body and the heat bath emit some non-self-interacting massless matter fields with Planckian distribution (we call these fields {\it the radiation fields} hereafter). Then hollow a spherical shell region out of the heat bath around the black body as shown in figure \ref{fig-1}. The outermost radius $R_h$ of hollow region satisfies $R_h \ge R_g$ by definition and the hollow region is concentric with the black body. The hollow region is filled up with the radiation fields emitted by the black body and the heat bath. Further we put two assumptions as follows.
\item[Fast propagation assumption:] The volume of hollow region is not so large that the speed of light is approximated as infinity. Then the retarded effect on radiation fields during propagating in the hollow region is ignored.
\item[Quasi-equilibrium assumption:] The time evolution in the NPT model is not so fast that thermodynamic states of black body and heat bath at each moment of their evolution are approximated well by equilibrium states individually. (Recall a quasi-static process in ordinary thermodynamics.) This is consistent with equations of states \eqref{eq-eq.eos} which is obtained from a static black hole solution of the Einstein equation (Schwarzschild black hole). Further, since a Schwarzschild black hole is not a quantum one, the order relation $R_g > O(1)$ should be required as well.
\end{description}
Here note that, as mentioned in equation \eqref{eq-eq.capacity}, the energetic and entropic properties of a Schwarzschild black hole are encoded in equations of states \eqref{eq-eq.eos}. Henceforth, as done in reference \cite{ref-trans.1}, we call the black body {\it the black hole}. The reason why we consider an isolated case of the whole system is that we put our basis on the model used in section \ref{sec-eq} and reference \cite{ref-trans.1}, as mentioned at the end of section \ref{sec-eq}.

There are three points which have to be mentioned here. The first point is about the thermodynamic states of black hole and heat bath under the quasi-equilibrium assumption. This denotes that time evolution of black hole is expressed as that thermodynamic state of black hole changes along a sequence of equilibrium states in the state space during time evolution of the whole system. And the same is true of time evolution of thermodynamic state of heat bath. Then the temperatures $T_g$ and $T_h$ are given as equilibrium temperatures. Therefore it is reasonable to consider that, during the time evolution, the temperatures $T_g$ and $T_h$ are constant within a time scale that one particle of radiation fields travels in the hollow region until absorbed by black hole or heat bath. This is consistent with the fast propagation assumption.

The second point is about the thermodynamic state of radiation fields under the quasi-equilibrium assumption. In the hollow region, the radiation fields of different temperatures $T_g$ and $T_h$ are simply superposed, since the radiation fields are non-self-interacting (a gas of collisionless particles). That is, the radiation fields are in a two-temperature nonequilibrium state. Here note that, since the temperatures $T_g$ and $T_h$ are constant within the time scale that one particle of radiation fields travels in the hollow region, it is reasonable to consider that the radiation fields have a stationary energy flow from the black hole to the heat bath (or from the heat bath to the black hole) within that time scale. Hence, at each moment of time evolution of the whole system, the thermodynamic state of radiation fields is well approximated to be a macroscopically stationary nonequilibrium state, which we call the {\it steady state} hereafter. The time evolution of radiation fields is expressed as that the thermodynamic state of radiation fields changes along a sequence of steady states in the state space. Therefore we need a thermodynamic formalism of two-temperature steady states for radiation fields in order to analyse the black hole phase transition and evaporation in the framework of NPT model. The steady state thermodynamics for radiation fields has already been formulated in reference \cite{ref-sst}, and its brief summary is in appendix \ref{app-sst}. We will make use of the steady state thermodynamics in analysing the NPT model.

The third point which should be mentioned here is about the value of $N$ which appears in the generalised Stefan-Boltzmann constant $\sigmap = N \pi^2/120$. As mentioned at the end of section \ref{sec-intro} and appendix \ref{app-sst}, $N$ denotes the number of independent states in radiation fields. Further, because of the quasi-equilibrium assumption, the NPT model should be valid at most until the semi-classical stage of black hole evaporation ends ($T_g <$ Planck temperature) and no exotic particle is created in the radiation fields by the Hawking effect. Therefore it is reasonable to estimate the value of $N$ by the number of independent states of standard particles (three generations of quarks and leptons and their internal states, and gauge particles of four fundamental interactions and their helicities). That is, we require the following condition hereafter,
\sikib
 N = O(10) \quad \left( \Rightarrow \,\,\, \sigmap = O(1) \, \right) \, .
\label{eq-ne.N}
\sikie

\subsection{Energy transport and the phase transition in NPT model}
\label{sec-ne.energy}

We discuss energetics of NPT model. The total energy of the whole system is $E_{tot} = E_g + E_h + E_{rad}$, where $E_g$ is the energy of black hole given in equations of states \eqref{eq-eq.eos}, $E_h$ is the energy of heat bath defined by ordinary thermodynamics, and $E_{rad}$ is the steady state energy of radiation fields in the hollow region. The energy $E_{rad}$ is given by reference \cite{ref-sst} or equations \eqref{eq-sst.energy.rad} and \eqref{eq-sst.fermion} in appendix \ref{app-sst} (the generalised Stefan-Boltzmann constant $\sigmap$ appears in the form of $E_{rad}$, since the radiation fields are non-self-interacting massless matter fields),
\sikib
 E_{rad} = 4 \sigmap \left( G_g \, T_g^4 + G_h \, T_h^4 \right) \quad , \quad
 G_g = \int_{V_{rad}} dx^3\, \gb{x} \quad , \quad G_h = \int_{V_{rad}} dx^3\, \gh{x} \, ,
\sikie
where $\gb{x}$ is the solid-angle (divided by $4 \pi$) covered by directions of particles which are emitted by the black hole and come to a point $\vec{x}$ (see figure \ref{fig-1} and figure \ref{fig-app.1} in appendix \ref{app-sst}), and $\gh{x}$ is defined similarly by particles emitted by the heat bath. Note that $\gb{x} + \gh{x} = 1$ holds by definition, and consequently $V_{rad} = G_g + G_h$ is the volume of hollow region. Further, since the black hole is concentric with the hollow region, we obtain $\gb{x} = \left(\, 1 - \cos\theta_r \,\right)/2$ and $\gh{x} = \left(\, 1 + \cos\theta_r \,\right)/2$, where $\theta_r$ is the zenith angle which covers the black hole at a point of radial distance $r$ (see right panel in figure \ref{fig-1}). Then the quantities $G_g$ and $G_h$ are calculated,
\sikib
 G_g = \frac{2\,\pi}{3}
       \left[\, R_h^3 - R_g^3 - \left(\, R_h^2 - R_g^2 \, \right)^{3/2} \, \right]
 \quad , \quad
 G_h = \frac{2\,\pi}{3}
       \left[\, R_h^3 - R_g^3 + \left(\, R_h^2 - R_g^2 \, \right)^{3/2} \, \right] \, .
\label{eq-ne.G}
\sikie

In order to understand the energy flow in NPT model, we divide the whole system into two sub-systems X and Y as follows. The sub-system X is composed of a black hole and ``out-going'' radiation fields emitted by the black hole, and the sub-system Y is composed of a heat bath and ``in-going'' radiation fields emitted by the heat bath (see left panel in figure \ref{fig-1}). That is, the sub-system X is a combined system of components of NPT model which share the temperature $T_g$, and Y is that which share the temperature $T_h$. Then the total energy of the whole system is expressed as
\sikib
 E_{tot} = E_X + E_Y \, ,
\label{eq-ne.total.energy}
\sikie
where $E_X$ and $E_Y$ are respectively the energies of sub-systems X and Y, 
\sikib
 \begin{cases}
  E_X = E_g + E_{rad}^{(g)} &,\quad E_{rad}^{(g)} = 4 \sigmap \, G_g \, T_g^4 \\
  E_Y = E_h + E_{rad}^{(h)} &,\quad E_{rad}^{(h)} = 4 \sigmap \, G_h \, T_h^4
 \end{cases} \, ,
\label{eq-ne.energyXY}
\sikie
where $E_{rad}^{(g)}$ and $E_{rad}^{(h)}$ are respectively the energies of out-going and in-going radiation fields. The energy flow in NPT model can be understood as an energy transport between sub-systems X and Y. In order to express the energy transport explicitly, recall that the energy exchange between a back hole and a heat bath is carried by the exchange of radiation fields emitted by them, as mentioned at the beginning of section \ref{sec-ne.model}. Hence the Stefan-Boltzmann law works well as an explicit expression of energy transport (with the generalised Stefan-Boltzmann constant $\sigmap$),
\sikib
 \frac{dE_X}{dt} = - \sigmap \left( T_g^4 - T_h^4 \right) A_g \quad , \quad
 \frac{dE_Y}{dt} = \sigmap \left( T_g^4 - T_h^4 \right) A_g \, ,
\label{eq-ne.transport.1}
\sikie
where $A_g = 4 \pi R_g^2$ is the surface area of black hole, and $t$ is a time coordinate which corresponds to a proper time of a rest observer at asymptotically flat region if we can extend the NPT model to a full general relativistic model. Note that, some of particles of the radiation fields emitted by heat bath are not absorbed by the black hole but return to the heat bath (see right panel in figure \ref{fig-1}). Therefore the effective surface area through which the sub-system Y exchanges energy with the sub-system X is equal to the surface area of black hole $A_g$, and the area $A_g$ appears in both of equations \eqref{eq-ne.transport.1}. Further we note that the energy transport \eqref{eq-ne.transport.1} is formulated to be consistent with the isolated setting of the NPT model ($E_{tot} = constant$).

It is useful to rewrite the energy transport \eqref{eq-ne.transport.1} to a more convenient form for later discussions. By equations \eqref{eq-eq.capacity}, \eqref{eq-ne.G} and \eqref{eq-ne.energyXY}, the energy transport \eqref{eq-ne.transport.1} becomes
\sikib
 C_X\,\frac{dT_g}{dt} = - J \quad , \quad
 C_X\,C_Y^{(h)}\,\frac{dT_h}{dt} = \left(\,C_X + C_Y^{(g)} \,\right)\,J \, ,
\label{eq-ne.transport.2}
\sikie
where
\sikib
 J = \sigmap \left(\, T_g^4 - T_h^4 \,\right) A_g \, ,
\label{eq-ne.J}
\sikie
and
\sikib
 \begin{cases}
 C_X = \dfrac{dE_X}{dT_g} = C_g + C_{rad}^{(g)} &, \quad
 C_{rad}^{(g)} = \dfrac{dE_{rad}^{(g)}}{dT_g}
               = 16\,\sigmap\,G_g\,T_g^3
                + \dfrac{\sigmap}{2\,\pi}\left(R_g - \sqrt{R_h^2 - R_g^2} \right)T_g \\
 C_Y^{(g)} = \dfrac{\partial E_Y}{\partial T_g}
           = \dfrac{\sigmap}{2\,\pi}\left(R_g + \sqrt{R_h^2 - R_g^2} \right)
                                    \dfrac{T_h^4}{T_g^3} &\\
 C_Y^{(h)} = \dfrac{\partial E_Y}{\partial T_h}
           = C_h + 16\,\sigmap\,G_h\,T_h^3 &, \quad
 C_h = \dfrac{dE_h}{dT_h} \, ( > 0)
 \end{cases}
\label{eq-ne.capacity}
\sikie
where $C_g = - 2 \pi R_g^2$ is given by equation \eqref{eq-eq.capacity} and it is assumed for simplicity that $R_h = constant$ and that $E_h$ depends on $T_h$ but not on $T_g$. The quantity $C_h$ is the heat capacity of heat bath, and we assume $C_h = constant > 0$ for simplicity. The quantity $C_Y^{(g)}$ is the heat capacity of sub-system Y under the change of $T_g$ with fixing $T_h$, and $C_Y^{(h)}$ is that under the change of $T_h$ with fixing $T_g$. The quantity $C_{rad}^{(g)}$ is the heat capacity of out-going radiation fields, and $C_X$ is the heat capacity of sub-system X. Here note that $E_X$ has no $T_h$-dependence. In analysing the nonlinear differential equations \eqref{eq-ne.transport.2}, behaviours of various heat capacities \eqref{eq-ne.capacity} are used. Some useful properties of these heat capacities are explained in appendices \ref{app-CX.Rg}, \ref{app-CX.Rh}, \ref{app-Cg/CX} and \ref{app-CXCYg}.

Here it should be pointed out that an inequality $C_X + C_Y^{(g)} < 0$ has to hold in order to guarantee the validity of NPT model. To understand this requirement, consider the case $T_g > T_h$ for the first. The energy flows from a black hole to a heat bath via radiation fields, $dE_g < 0$ and $dE_h > 0$. Then $dT_g > 0$ and $dT_h > 0$ hold due to $C_g < 0$ and $C_h > 0$. Recall that we are considering an isolated case of the whole system, $E_{tot} = constant \, \Rightarrow \, ( C_X + C_Y^{(g)} ) \,dT_g + C_Y^{(h)}\,dT_h = 0$. Therefore, because of $C_Y^{(h)} > 0$ by definition, it is concluded that the inequality $C_X + C_Y^{(g)} < 0$ must hold. And an inequality $C_X < 0$ follows immediately due to $C_Y^{(g)} > 0$ by definition. The similar discussion holds for the case $T_g < T_h$, and gives the same inequality. Therefore the following inequality must hold in the framework of NPT model,
\sikib
 C_X + C_Y^{(g)} < 0 \quad ( \, \Rightarrow \,C_X < 0 \, )  \, .
\label{eq-ne.validity.1}
\sikie
This inequality is the condition which guarantees the validity of NPT model. A more detailed property of the combined heat capacity $C_X + C_Y^{(g)}$ is explained in appendix \ref{app-CXCYg}, which shows that inequality \eqref{eq-ne.validity.1} can hold for a sufficiently small $T_h$. Therefore we assume that $T_h$ is small enough so that this validity condition \eqref{eq-ne.validity.1} holds.

Concerning the validity of NPT model, it is also important to consider what the quasi-equilibrium assumption implies. This assumption requires that the time evolution is not so fast. Therefore it should be required that, either when a black hole evaporates ($T_g > T_h$) or when it grows ($T_g < T_h$), the shrinkage (or expansion) speed of the surface of black hole is less than unity, $\left| dR_g/dt \right| < 1$. Using equations \eqref{eq-ne.transport.2}, this becomes
\sikib
 \left| \dfrac{d R_g}{dt} \right| < 1 \quad \Rightarrow \quad
 \frac{1}{4 \pi} \left| \frac{J}{T_g^2 \, C_X} \right| < 1 \, .
\label{eq-ne.validity.2}
\sikie
This inequality is also the condition which guarantees the validity of NPT model. Hence, in the framework of NPT model, the analysis should be restricted within the situations which satisfy the conditions \eqref{eq-ne.validity.1} and \eqref{eq-ne.validity.2}.

Here it is helpful for later discussions to consider what a violation of validity conditions \eqref{eq-ne.validity.1} and \eqref{eq-ne.validity.2} denotes. Firstly consider if the validity condition \eqref{eq-ne.validity.1} is not satisfied. Then the system, especially the radiation fields, can never be described with steady state thermodynamics (``steady'' means ``stationary nonequilibrium''). That is, the radiation fields should be in a highly nonequilibrium dynamical state. Then the quasi-equilibrium assumption is violated, because it is this assumption that lead us to utilise the steady state thermodynamics as explained in section \ref{sec-ne.model}. Therefore highly nonequilibrium radiation fields make a black hole dynamical, and the black hole can not be treated by a stationary solution of the Einstein equation. Next consider if the validity condition \eqref{eq-ne.validity.2} is not satisfied. Then a black hole evolves so fast that the quasi-equilibrium assumption is violated and that the black hole can not be described by a stationary solution of the Einstein equation. That is, a black hole should be described as a dynamical one, and consequently radiation fields evolve into a highly nonequilibrium dynamical state. Hence it is expected that, when one of the validity conditions \eqref{eq-ne.validity.1} or \eqref{eq-ne.validity.2} is violated, the system evolves into a highly nonequilibrium dynamical state which can not be treated in the framework of NPT model.

Finally in this section, we discuss about the black hole phase transition in the NPT model. If the whole system is in an exact equilibrium, then a black hole, a heat bath and radiation fields have the same equilibrium temperature $T_{eq} \,(= T_g = T_h)$, and, as shown in section \ref{sec-eq}, one may think that an instability of equilibrium occurs when an inequality $\left| C_g \right| < C_h^{(eq)}$ holds, where $C_h^{(eq)} = C_h + 16\,\sigmap\,T_{eq}^3\,V_{rad}$ and $16\,\sigmap\,T_{eq}^3\,V_{rad}$ is the heat capacity of radiation fields in an equilibrium. However we should be more careful to consider the criterion of occurrence of instability of the whole equilibrium state in the framework of NPT model. At the occurrence of instability, the temperature of black hole $T_g$ and that of heat bath $T_h$ become different from each other due to the temperature fluctuation. Then, referring to the conservation of total energy, $dE_{tot} = ( C_X + C_Y^{(g)} ) \,dT_g + C_Y^{(h)}\,dT_h = 0$, a relation $dT_g > dT_h$ holds when the following inequality is satisfied with the fluctuated temperatures $T_g$ and $T_h \,( \neq T_g )$,
\sikib
 \left| C_X + C_Y^{(g)} \right| < C_Y^{(h)} \quad \Rightarrow \quad
 \left| C_g \right| < C_{rad}^{(g)} + C_Y^{(g)} + C_Y^{(h)} \, ,
\label{eq-ne.criterion}
\sikie
where $C_{rad}^{(g)} > 0$ is used which is shown in appendix \ref{app-Cg/CX}. This inequality denotes a growth of temperature difference $\delta T = T_g - T_h$. That is, we can expect that the criterion of occurrence of instability is given by inequality \eqref{eq-ne.criterion}. However it should be noted that the temperatures $T_g$ and $T_h$ evolve in time for the case $T_g \neq T_h$. Therefore, even if inequality \eqref{eq-ne.criterion} is not satisfied with specific values of fluctuated temperatures $T_g$ and $T_h$ at an initial time, there is a possibility that this inequality comes to be satisfied and an instability occurs during time evolution of the system (an example is shown in next section). And a converse situation is also possible that this inequality comes not to be satisfied even if it is satisfied at an initial time. Hence, when only initial values of fluctuated temperatures $T_g$ and $T_h$ are used, we can not exactly judge by inequality \eqref{eq-ne.criterion} whether an instability of the whole equilibrium state occurs or not. That is, in a practical usage, inequality \eqref{eq-ne.criterion} is not useful as a criterion of occurrence of instability. Here it should be noted as well that, contrary to the NPT model, the criterion of occurrence of instability in the equilibrium model used in section \ref{sec-eq} is given by the inequality $\left| C_g \right| < C^{(eq)}$ evaluated at the initial time. Hence the nonequilibrium effect of energy exchange tends to make the situation more complicated.

\subsection{Black hole evaporation in a heat bath}
\label{sec-ne.evapo}

\subsubsection{General aspect of the NPT model}

In order to discuss a black hole evaporation after the instability of an equilibrium of the whole system occurs, we assume $T_g > T_h$ hereafter. In order to analyse energy transport equations \eqref{eq-ne.transport.2} from energetic viewpoint, we consider the energy emission rate $\jn$ by black hole in the framework of NPT model,
\sikib
 \jn = - \frac{dE_g}{dt}
  = \sigmap \, \frac{C_g}{C_X} \, \left(T_g^4 - T_h^4\right) \, A_g \, ,
\label{eq-ne.rate.npt}
\sikie
where equations \eqref{eq-eq.capacity} and \eqref{eq-ne.transport.2} are used. The larger the value of $\jn$, the more rapidly the energy $E_g$ of black hole decreases along its evaporation process. In other words, the stronger emission rate $J_{npt}$ denotes the acceleration of black hole evaporation.

Here recall that, as mentioned in equations \eqref{eq-ne.capacity}, we assume $R_h = constant$ for simplicity as the evaporation process proceeds. However, because $R_h$ is the parameter which controls the size of nonequilibrium region around black hole, it is useful to compare two situations which differ only by the value of $R_h$ with sharing the same values of the other parameters of NPT model, $R_g$, $T_h$, $C_h$ and $N$ in $\sigmap$. In order to make this comparison, we note the following three points; firstly $C_g < 0$ by definition, secondly $C_X < 0$ shown by inequality \eqref{eq-ne.validity.1}, and finally that $\left| C_X \right|$ is monotone decreasing as a function of $R_h$ for $R_h \ge \left(3\sqrt{2}/4\right) R_g \simeq 1.06 R_g$ (see appendix \ref{app-CX.Rh}). The first and second points denote $C_g/C_X > 0$, then the third point concludes that $C_g/C_X$ is monotone increasing as a function of $R_h$ for $R_h \ge \left(3\sqrt{2}/4\right) R_g$. Hence it is recognised that, for the case of $R_h > \left(3\sqrt{2}/4\right) R_g$, the larger we set the nonequilibrium region, the faster the black hole evaporation process proceeds. Numerical examples are shown later in figure \ref{fig-2}.

The above discussion is a comparison of NPT model of a certain value of $R_h$ with that of a different value of $R_h$. In the following subsections, we compare the NPT model with the other models of black hole evaporation, the equilibrium model used in section \ref{sec-eq} and the black hole evaporation in an empty space (a situation without heat bath).

\subsubsection{Comparison with the equilibrium model used in section \ref{sec-eq}}

Since the NPT model is constructed from the equilibrium model used in section \ref{sec-eq}, it is interesting to compare the NPT model with the equilibrium model. The energy emission rate $J_{eq}$ by black hole in the equilibrium model is given with setting $R_h = R_g$ in $\jn$ for all time,
\sikib
 J_{eq} = \sigmap \left(T_g^4 - T_h^4\right) A_g = J \, ,
\sikie
where $J$ is given by equation \eqref{eq-ne.J}. Then we find $\jn = \left( C_g/C_X \right) \, J_{eq}$. Note that appendix \ref{app-Cg/CX} shows $C_g/C_X > 1$. Therefore, when the values of $R_g$, $T_h$ and $N$ in $\sigmap$ are the same for the NPT and equilibrium models, then $\jn > J_{eq}$ holds. This implies that the black hole evaporation in NPT model proceeds faster than that in the equilibrium model. That is, we can recognise that the nonequilibrium effect of energy exchange between a black hole and a heat bath accelerates the black hole evaporation.

\subsubsection{Comparison with the black hole evaporation in an empty space}

Usually in many papers on black hole physics, the evaporation time scale of black hole is estimated with assuming that a black hole is in an empty space (a situation without heat bath). It is useful to compare the NPT model with the black hole evaporation in an empty space. When gravitational interactions between a black hole and matter fields of Hawking radiation are ignored (see appendix \ref{app-essence}), the energy emission rate $\je$ by black hole in an empty space is given by simple Stefan-Boltzmann law in an empty space ($dE_g/dt = - \sigmap T_g^4 A_g$),
\sikib
 \je = \sigmap \, T_g^4 \, A_g \, ,
\label{eq-ne.rate.empty}
\sikie
where it is assumed that matter fields of Hawking radiation is non-self-interacting massless matter fields (this is the same ``radiation fields'' as those considered in the NPT model). Then we find
\sikib
 \jn = \frac{C_g}{C_X} \left( 1 - \frac{T_h^4}{T_g^4} \right) \je \, .
\label{eq-ne.jn.je}
\sikie
Recall that $T_g > T_h \,( \Rightarrow \, 1 -T_h^4/T_g^4 < 1 )$ holds generally for a black hole evaporation, and $C_g/C_X > 1$ holds in the framework of NPT model (see appendix \ref{app-Cg/CX}). Then the factor $\left(C_g/C_X\right) \left( 1 - T_h^4/T_g^4 \right)$ may be greater or less than unity. Therefore it is not definitely clear which of $\jn$ and $\je$ is larger than the other.

One may naively expect that the incoming energy flow from heat bath to black hole in NPT model never enhance the energy emission rate by black hole, and that the relation $\jn > \je$ is impossible but $\jn < \je$ must hold always. It is always true if the out-going energy flow due to Hawking radiation is never absorbed by the heat bath. However in the NPT model, the energy emitted by Hawking radiation is absorbed by the heat bath and affects the incoming energy flow from heat bath to black hole. Then it is reasonable to expect that the energetic interaction (energy exchange) between black hole and heat bath determines the energy emission rate $\jn$. When we take the energetic interaction into account, a relation $\jn > \je$ which is naively unexpected is also possible as discussed in the following paragraphs.

In order to discuss the energy emission rate $\jn$ under the effects of energy exchange between a black hole and a heat bath, it is useful to recall the decomposition of the whole system of NPT model into the sub-systems X and Y, as considered in section \ref{sec-ne.energy}. Further it should be noted here that, from energetic viewpoint, a black hole evaporation in an empty space can be thought of as a relaxation process of an isolated system in which the sub-system Y is removed from the NPT model and the sub-system X is isolated. Therefore, for the black hole evaporation in an empty space, the energy emission rate $\je$ is the energy transport just inside the sub-system X (from black hole to out-going radiation fields), and no energy flows out of the sub-system X. However the energy transport \eqref{eq-ne.transport.1} of NPT model is the energy exchange between sub-systems X and Y. That is, in the NPT model, the energy $E_X$ of sub-system X is extracted by the sub-system Y due to the temperature difference $T_g > T_h$, and energy flows from X to Y. The black hole evaporation in NPT model is essentially different from the black hole evaporation in an empty space due to the energetic interaction (energy exchange) between sub-systems X and Y. This difference may be understood by considering a limit of the energy transport \eqref{eq-ne.transport.1} as follows: one may expect that the energy emission equation of black hole evaporation in an empty space, $dE_g/dt = - \je$, should be obtained from equations \eqref{eq-ne.transport.1} by the limit operations, $T_h \to 0$, $E_h \to 0$ (remove the sub-system Y) and $R_h \to \infty$ (infinitely large volume of out-going radiation fields). However these operations transform equations \eqref{eq-ne.transport.1} into the set of equations, $dE_g/dt = - \je$ and $0 = \je$. This contradicts the ``evaporation'', since an unphysical result $E_g = constant\, (= \infty)$ is obtained. This implies that the black hole evaporation in an empty space can not be described as some limit situation of the NPT model. 

Hence, in addition to the naive expectation $\jn < \je$, the opposite expectation $\jn > \je$ may be expected due to the energetic interaction as follows: because the energy source of out-going radiation fields is the mass energy $E_g$ of black hole, it seems that, the more amount of energy $E_{rad}^{(g)}$ of out-going radiation fields is extracted by the sub-system Y, the more amount of mass energy $E_g$ of black hole should be radiated by the Hawking radiation. That is, when the energy $E_X$ of sub-system X is extracted by the sub-system Y during a black hole evaporation process in NPT model, then the energy emission process by black hole (the black hole evaporation) is accelerated due to the energy extraction by sub-system Y. This implies $\jn > \je$.

The above discussion can be supported by the following rough analysis. When a black hole evaporates in NPT model, the temperature difference $\delta T = T_g - T_h$ should grow infinitely, $\delta T \to \infty$ (recall the discussion in inequality \eqref{eq-ne.criterion}). Then, because of equation \eqref{eq-ne.jn.je} together with the facts $1 - \left(T_h/T_g\right)^4 \to 1$ (as $\delta T \to \infty$) and $C_g/C_X > 1$ (see appendix \ref{app-Cg/CX}), the larger the temperature difference $\delta T$, the larger the ratio $\jn/\je$. Hence for the case of black hole evaporation in the NPT model, it is expected that the relation $\jn > \je$ comes to be satisfied during the evaporation process even if the relation $\jn < \je$ holds at initial time. And, if this relation $\jn > \je$ holds for a sufficiently long time during the evaporation process, the evaporation time scale in NPT model can be shorter than that in an empty space. That is, it is possible that the black hole evaporation in NPT model proceeds faster than that in an empty space, where black holes of the same initial radius are considered in both cases. In next subsection, numerical examples which support this discussion are shown.

\subsubsection{Numerical example}

\begin{figure}[t]
 \begin{center}
  \includegraphics[height=37mm]{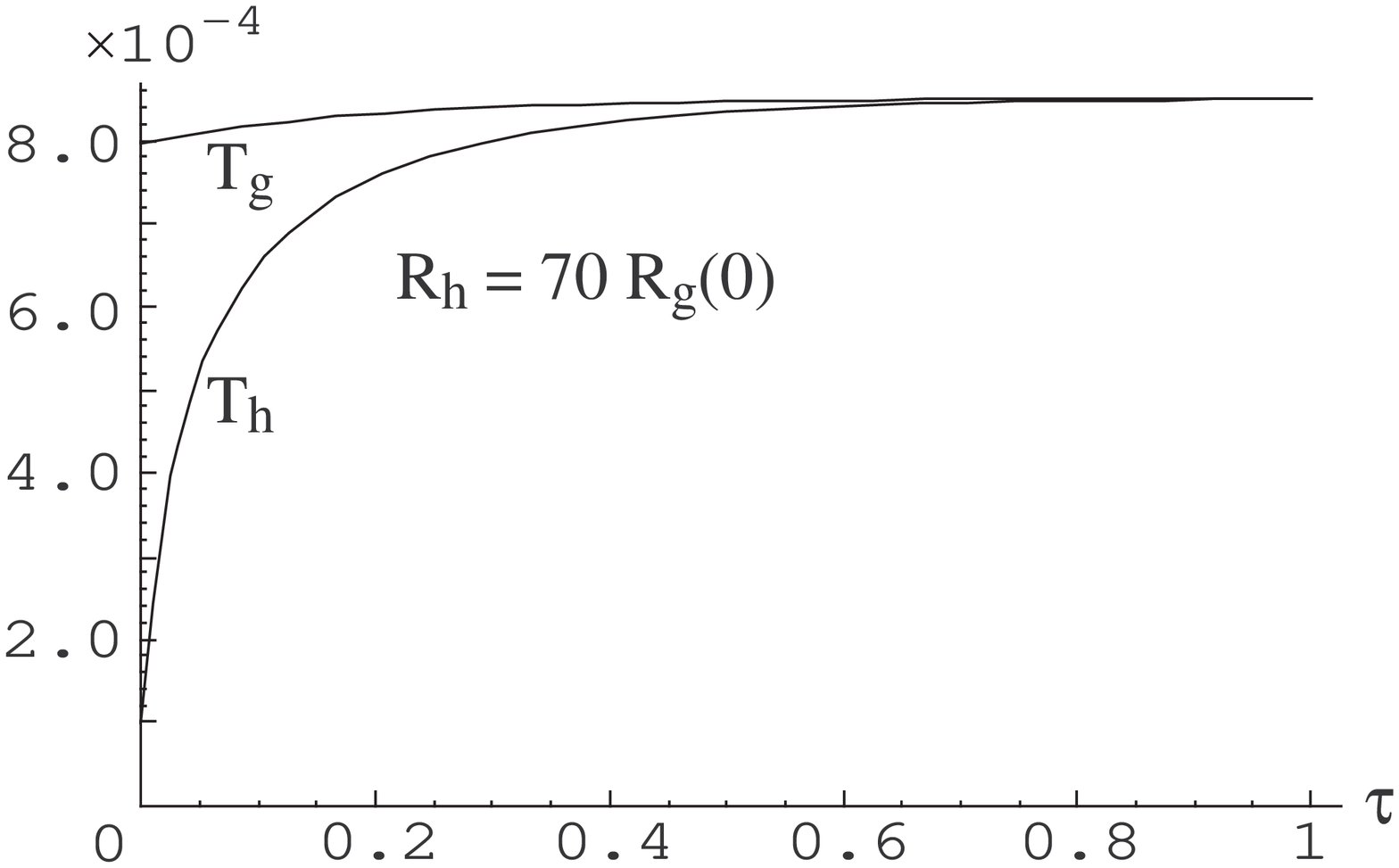}
  \includegraphics[height=37mm]{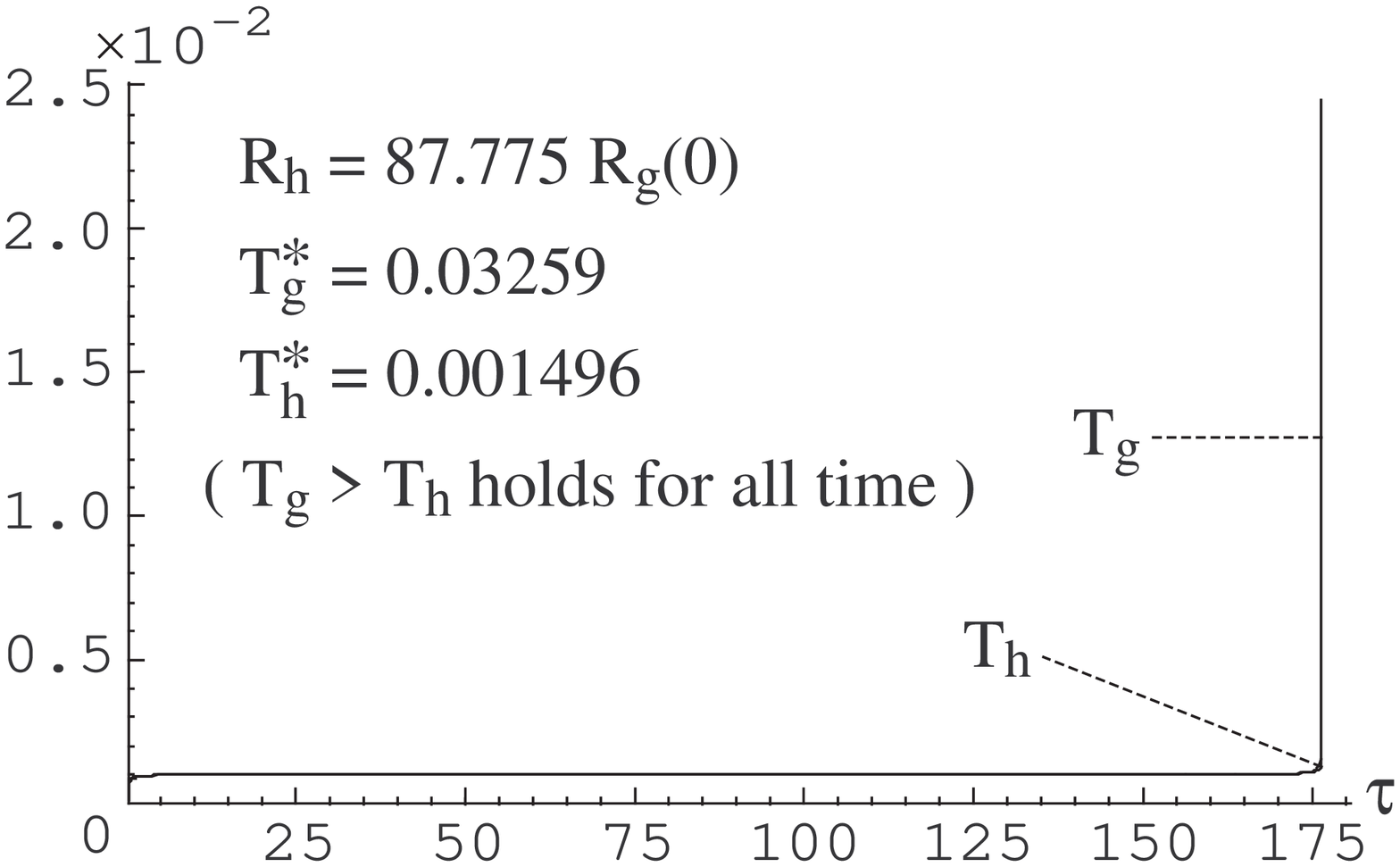}
  \includegraphics[height=37mm]{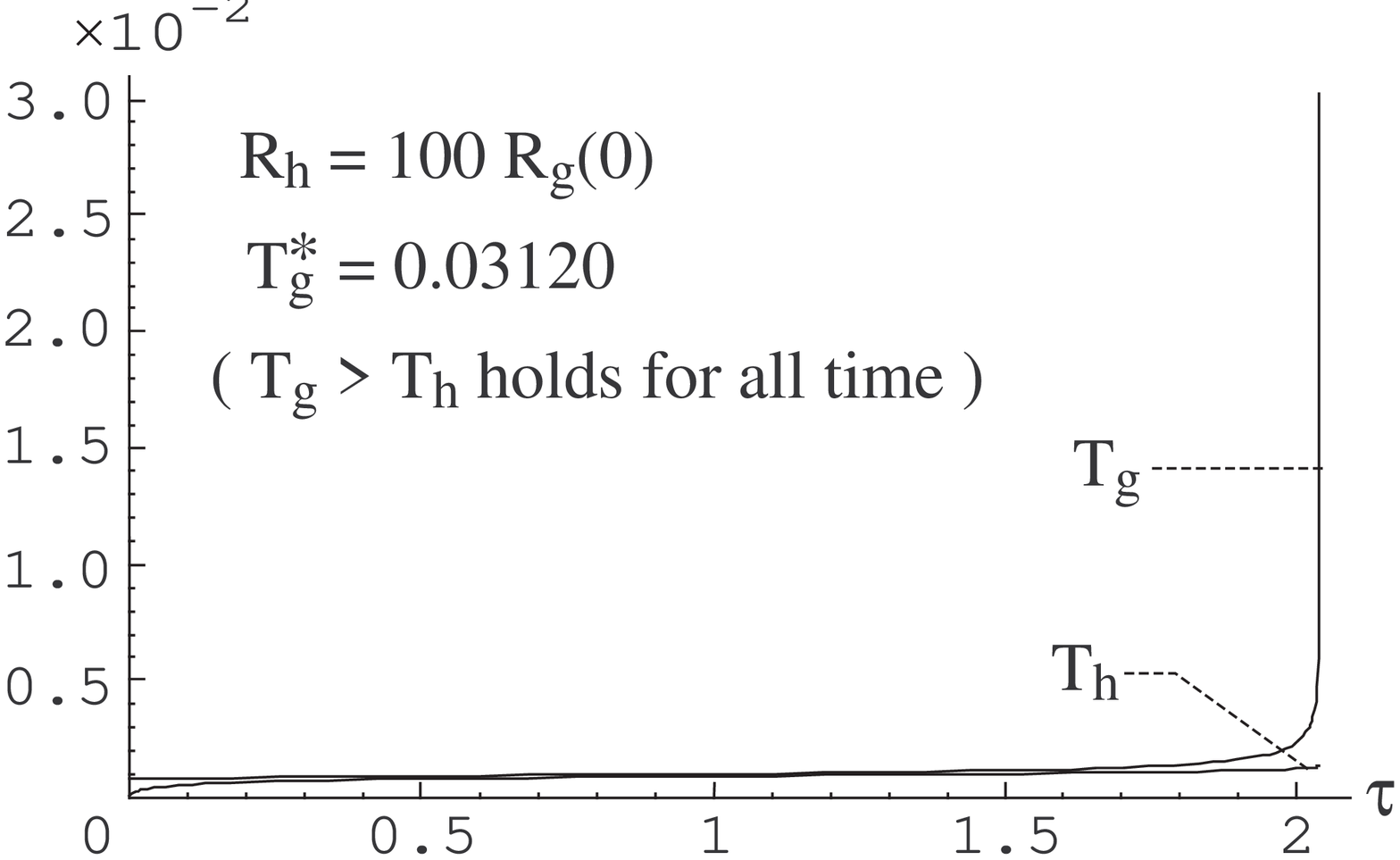} \\
  \includegraphics[height=37mm]{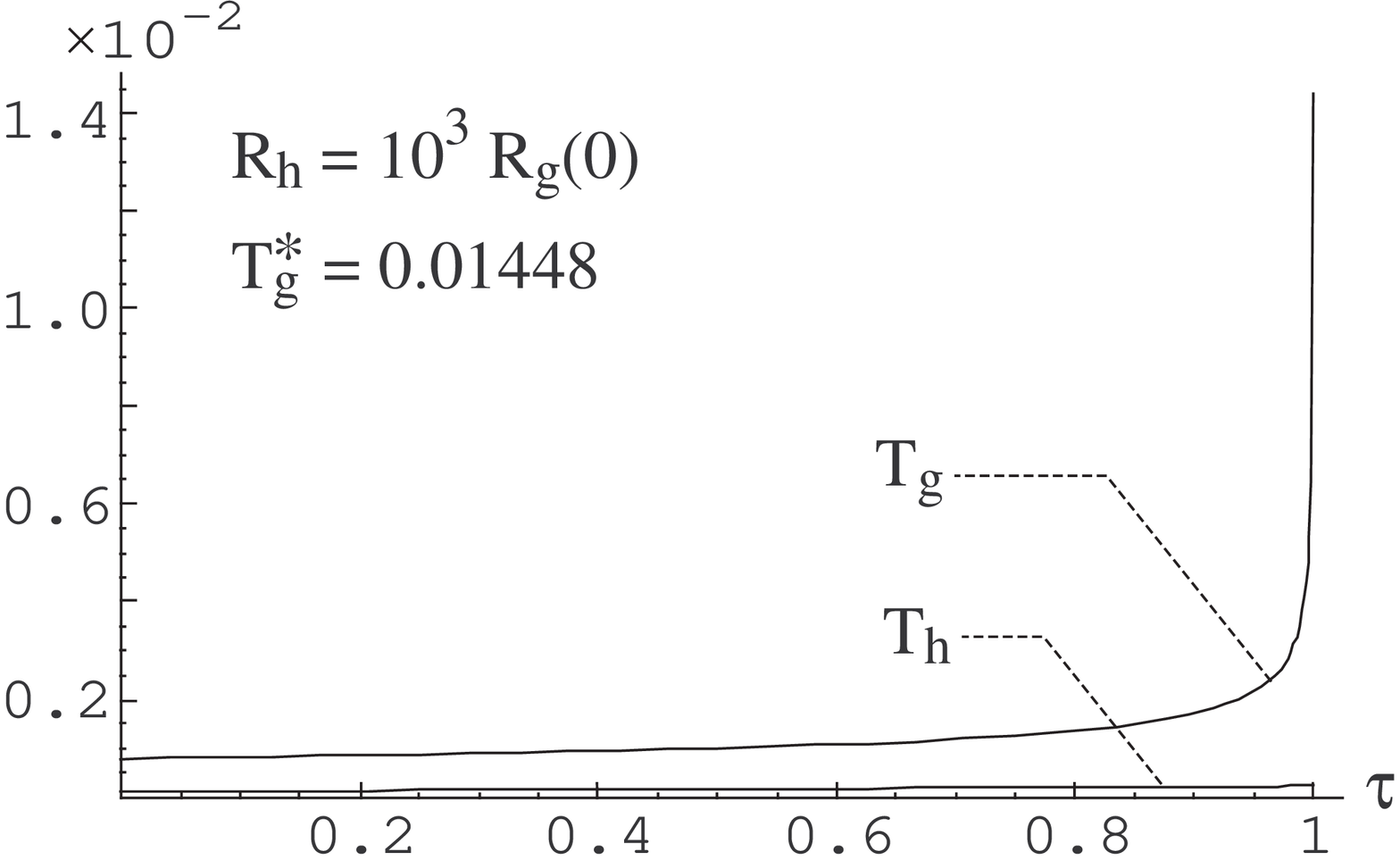}
  \includegraphics[height=37mm]{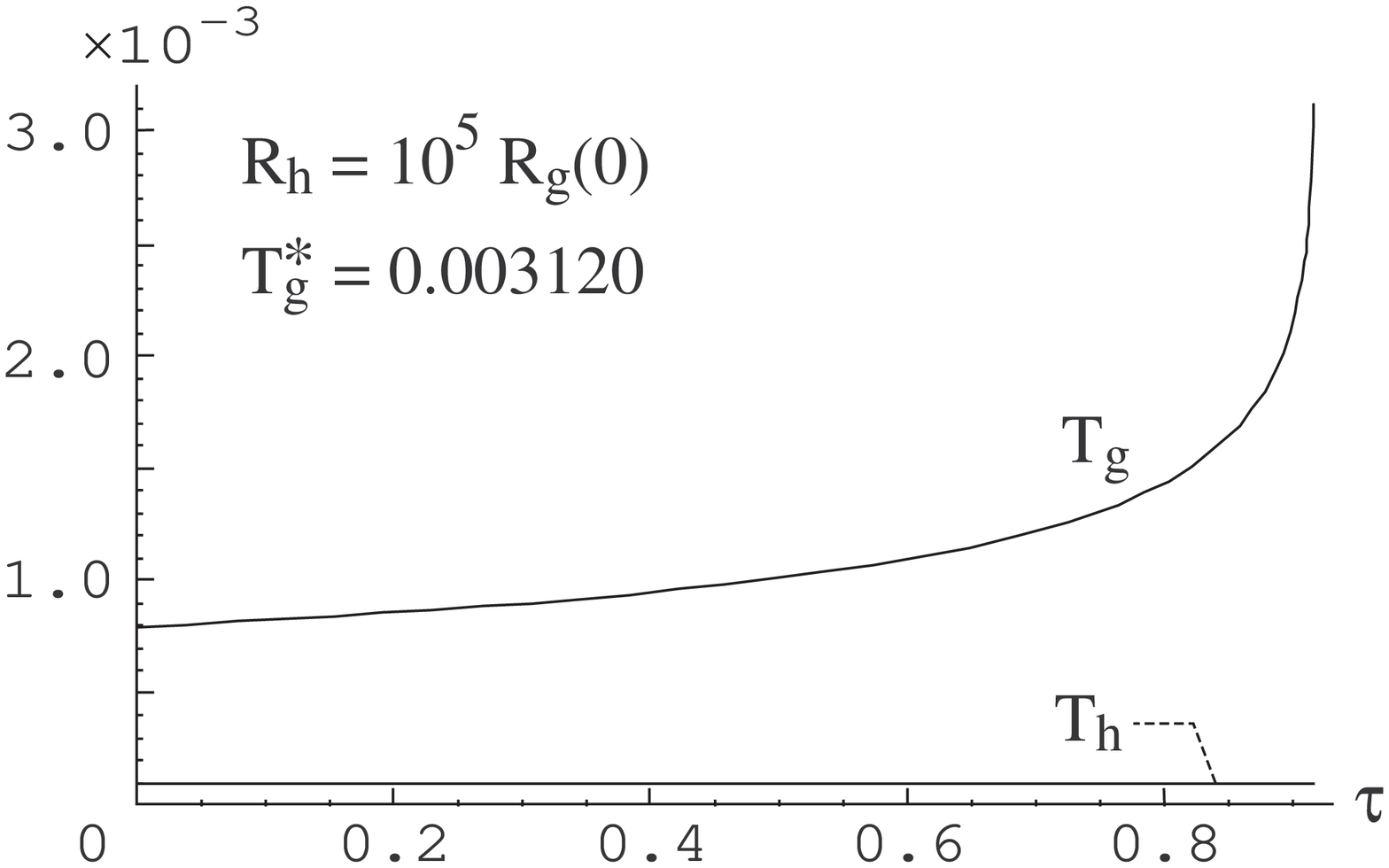}
  \includegraphics[height=37mm]{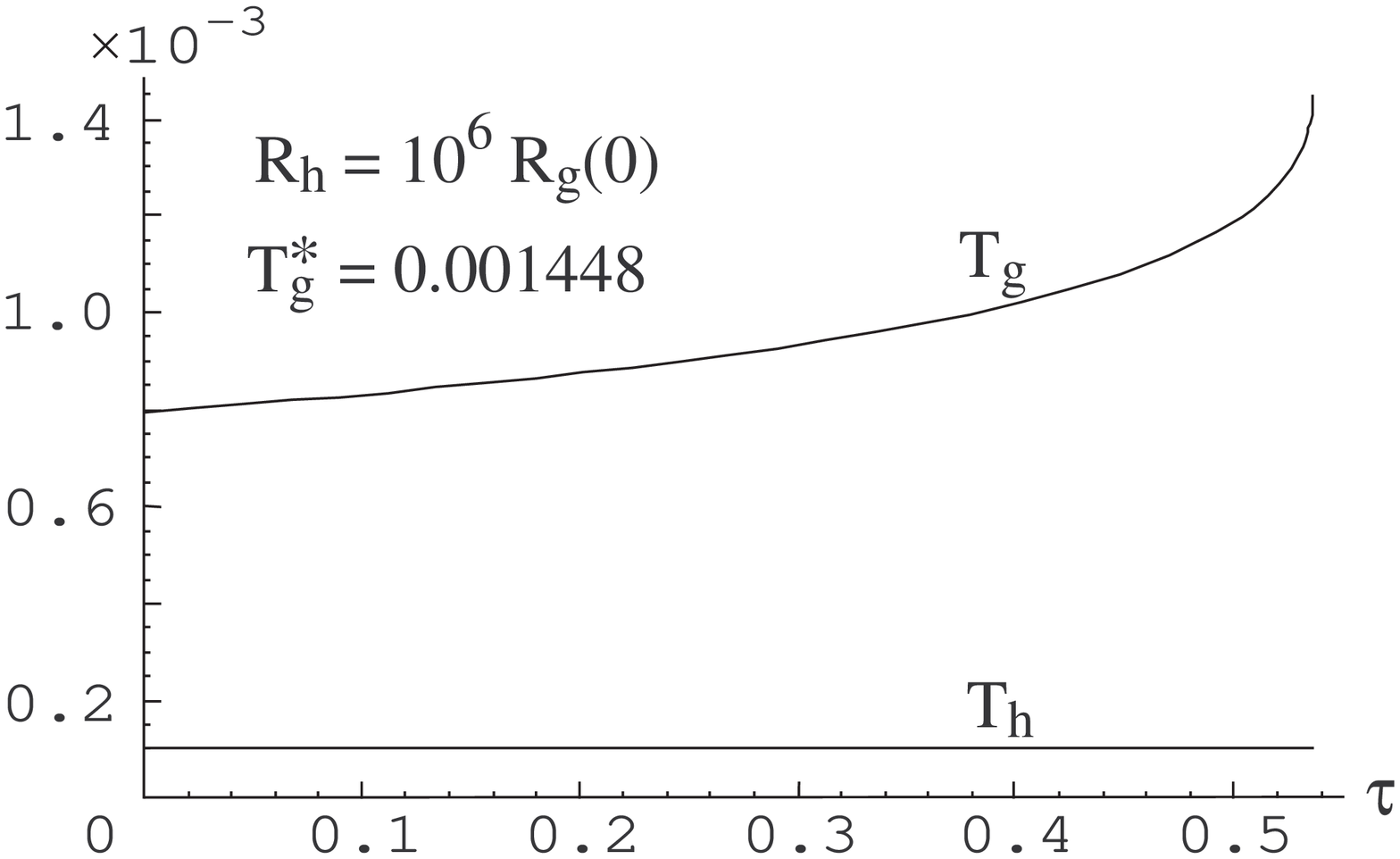}
 \end{center}
\caption{Numerical solutions of energy transport \eqref{eq-ne.transport.2} for $C_h = 10^3$, $N = 10$ and the initial conditions, $R_g(0) = 100$ and $T_h(0) = 0.0001$. Horizontal line denotes the normalised time $\tau = t/\te$. The size of nonequilibrium region is controlled by the outermost radius of hollow region $R_h$.}
\label{fig-2}
\end{figure}

Since an analysis of nonlinear differential equations is difficult, it is helpful to show numerical solutions $T_g(t)$ and $T_h(t)$ of energy transport equations \eqref{eq-ne.transport.2}. The initial conditions are
\sikib
 R_g(0) = 100 \,\, (\Rightarrow T_g(0) \simeq 0.00079) \quad , \quad T_h(0) = 0.0001 \, ,
\label{eq-ne.ic}
\sikie
and the other parameters are set as
\sikib
 C_h = 1000 \quad , \quad N = 10 \, ,
\label{eq-ne.Ch.N}
\sikie
where see equation \eqref{eq-ne.N} for the value of $N$. Further we have to specify the outermost radius $R_h$ of hollow region. As mentioned in equation \eqref{eq-ne.rate.npt}, by the comparison of a numerical solution of energy transport \eqref{eq-ne.transport.2} of a certain value of $R_h$ with that of a different value of $R_h$, we can observe the nonequilibrium effect of energy exchange between a black hole and a heat bath. The numerical results are shown in figure \ref{fig-2}, and the value of $R_h$ is attached in each panel. Time coordinate $\tau$ in this figure is a time normalised as
\sikib
 \tau = \dfrac{t}{\te} \, ,
\sikie
where $\te$ is the evaporation time (life time) of a black hole in an empty space (a situation without heat bath), which is determined as
\sikib
 \frac{dE_g}{dt} = -\je \quad \Rightarrow \quad
 R_g(t) = R_g(0)\, \left( 1 - \frac{N\, t}{1280 \, \pi\, R_g(0)^3} \right)^{1/3}
 \quad \Rightarrow \quad \te = \frac{1280 \, \pi}{N} \, R_g(0)^3 \, ,
\sikie
where one of equations \eqref{eq-eq.eos} is used, and $\te \simeq 4.02 \times 10^8$ by conditions \eqref{eq-ne.ic} and \eqref{eq-ne.Ch.N}. This $\te$ is usually adopted as the time scale of black hole evaporation in many papers on black hole physics. 

Further hereafter, we introduce the other time scale $\tn$ which denotes a time at which one of the validity conditions of NPT model \eqref{eq-ne.validity.1} or \eqref{eq-ne.validity.2} is broken,
\sikib
 \tn =
 \min\left[ \, t_1 , t_2 \, \left| \,
           C_X(t_1) + C_Y^{(g)}(t_1) = 0 \,,\, v(t_2) = 1 \right. \right] \, ,
\label{eq-ne.tnpt}
\sikie
where $C_X + C_Y^{(g)}$ is treated as a function of time $t$ through $R_g(t)$, and $v(t) = \left|dR_g(t)/dt\right|$. Each panel in figure \ref{fig-2} shows time evolutions of $T_g(t)$ and $T_h(t)$ for $0 < t < \tn$. The black hole temperature at $\tn$ is denoted as $T_g(\tn) = T_g^{\ast}$, which is also attached in each panel in figure \ref{fig-2}. The time $\tn$ obtained from our numerical results are listed;
\sikib
\begin{array}{|c||*{6}{c|}} \hline
  R_h/R_g(0) & 87.775  & 100   & 10^3   & 10^5   & 10^6   \\ \hline\hline
  \tn/\te    & 176.5   & 2.039 & 0.9989 & 0.9155 & 0.5363 \\ \hline
  R_g^{\ast} & 2.441   & 2.549 & 5.493  & 25.49  & 54.93  \\ \hline
  R_h/\tn    & 1.2 \times 10^{-7}
             & 1.2 \times 10^{-5}
             & 2.4 \times 10^{-4}
             & 2.7 \times 10^{-2}
             & 4.6 \times 10^{-1} \\ \hline
\end{array}
\label{eq-ne.list}
\sikie
For the cases of $R_h/R_g(0) = 87.775$, $100$, $10^3$ and $10^5$, the numerical plots stopped by $v(\tn) = 1$, but for the case of $R_h/R_g(0) = 10^6$, it stopped by $C_X(\tn) + C_Y^{(g)}(\tn) = 0$. This list (of $\tn$) supports the discussion given at the beginning of this section \ref{sec-ne.evapo} that the larger the radius $R_h$, the faster the black hole evaporation proceeds (the shorter the time $\tn$). The third line in this list is for $R_g^{\ast} = 1/(4 \pi T_g^{\ast})$, which gives an important information in next subsection. The lowest line shows that our numerical results are consistent with the fast propagation assumption. To understand this statement, consider a typical time scale $t_{rad}$ in which one particle of radiation fields travels in the hollow region from a black hole to a heat bath. The time scale $t_{rad}$ can be given by $t_{rad} = R_h$, since the radiation fields are massless. Then, if the time $t_{rad}$ is shorter than the time $\tn$, it is appropriate to consider that the fast propagation assumption is reasonable. Then the ratio $t_{rad}/\tn \, ( = R_h/\tn )$ shown at the lowest line in list \eqref{eq-ne.list} indicates that the fast propagation assumption is well satisfied.

If the nonequilibrium region is ignored and the equilibrium model used in section \ref{sec-eq} is considered with the same setting parameters, we find $\left| C_g \right| = 2\, \pi \times 100^2 > 1000 = C_h$, and the black hole settles down to a stable equilibrium state with a heat bath as explained in section \ref{sec-eq}. That is, the occurrence of accelerated increase of temperature $T_g$ in figure \ref{fig-2} is obviously due to the nonequilibrium effect of energy exchange between black hole and heat bath. For the case $R_h = 70 \, R_g(0)$, the black hole settles down to a stable equilibrium state, since $| C_X + C_Y^{(g)} | > C_Y^{(h)}$ holds for all time in this case (recall the discussion given in inequality \eqref{eq-ne.criterion}). As the radius $R_h$ is set larger, inequality \eqref{eq-ne.criterion} comes to be satisfied in the course of time evolution of the system. Then the black hole evaporation can be observed as an accelerated increase of temperature difference $T_g - T_h$ for the case $R_h = 87.775 \, R_g(0)$. The time scale $\tn$ of this case is very longer than the time $\te$. However, as the radius $R_h$ is set larger and the nonequilibrium region becomes larger, the time scale $\tn$ becomes shorter and we find $\tn \simeq \te$ about $R_h \simeq 10^3 \, R_g(0)$. And $\tn$ becomes shorter to $\tn \simeq 0.5363 \, \te$ at $R_h = 10^6 \, R_g(0)$. If we set $R_h \gtrsim 10^7$, the combined heat capacity becomes positive $C_X + C_Y^{(g)} > 0$ which is forbidden in the framework of NPT model as explained in inequality \eqref{eq-ne.validity.1}. That is, a black hole with nonequilibrium region of $R_h \gtrsim 10^7$ can not be treated by the NPT model, and, as discussed in section \ref{sec-ne.energy} (the paragraph one before the last one), a black hole evaporation for this case should be described as a highly nonequilibrium dynamical process in which the black hole can not be treated with a stationary black hole solutions of the Einstein equation. The larger the nonequilibrium region, the faster the black hole evaporation process evolves into a highly nonequilibrium dynamical stage.

\begin{figure}[t]
 \begin{center}
  \includegraphics[height=34mm]{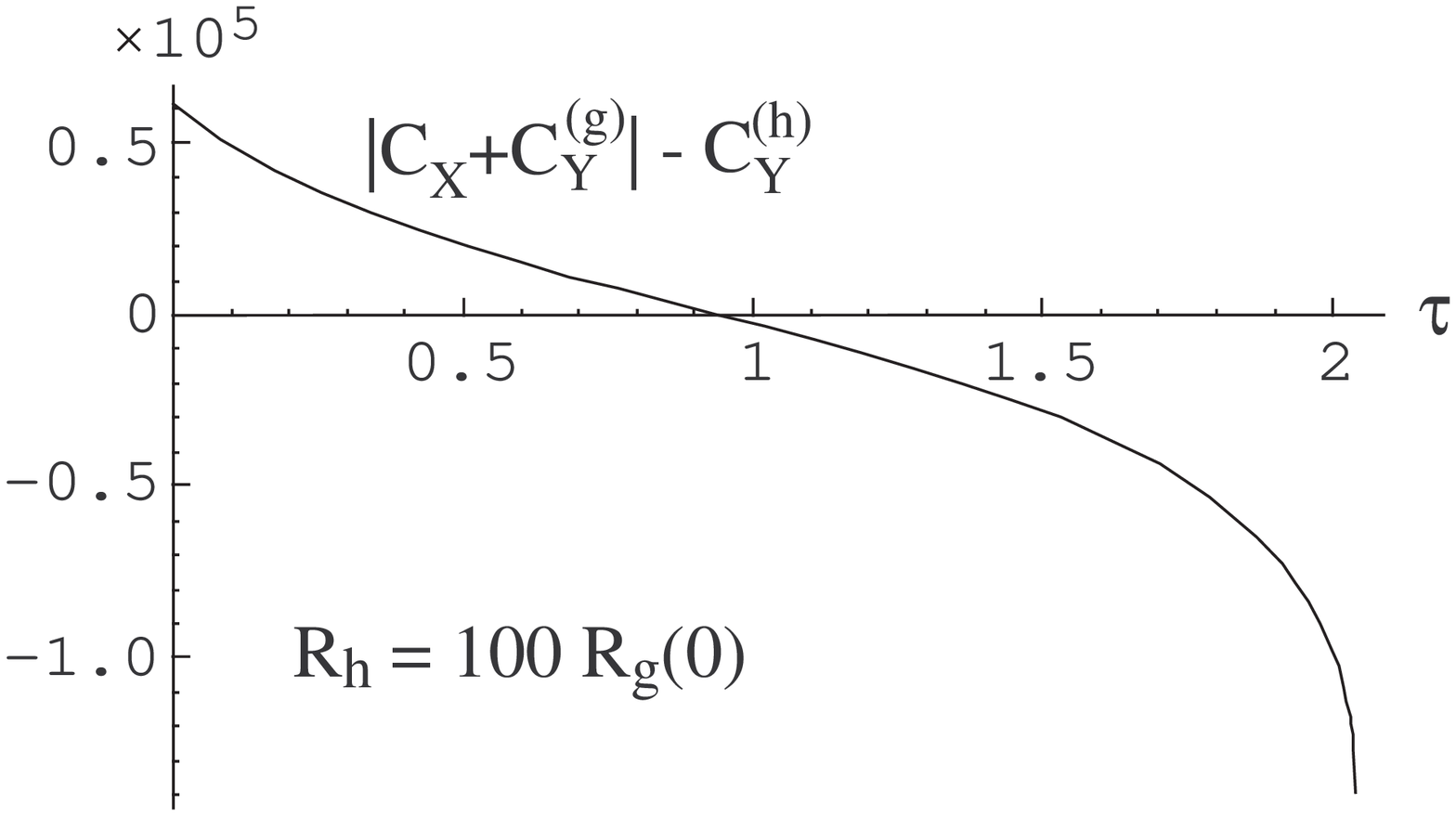}
  \includegraphics[height=34mm]{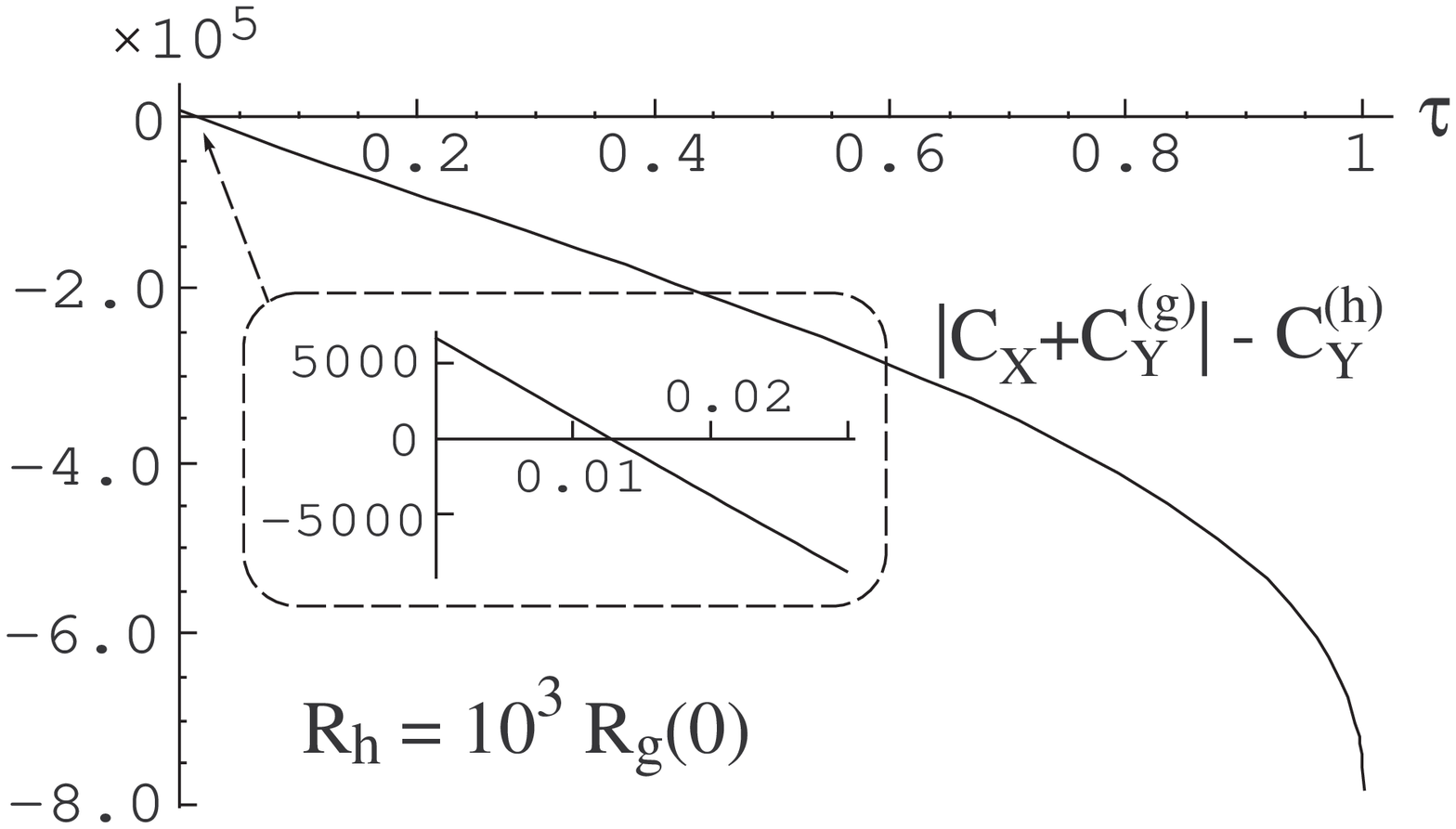}
  \includegraphics[height=34mm]{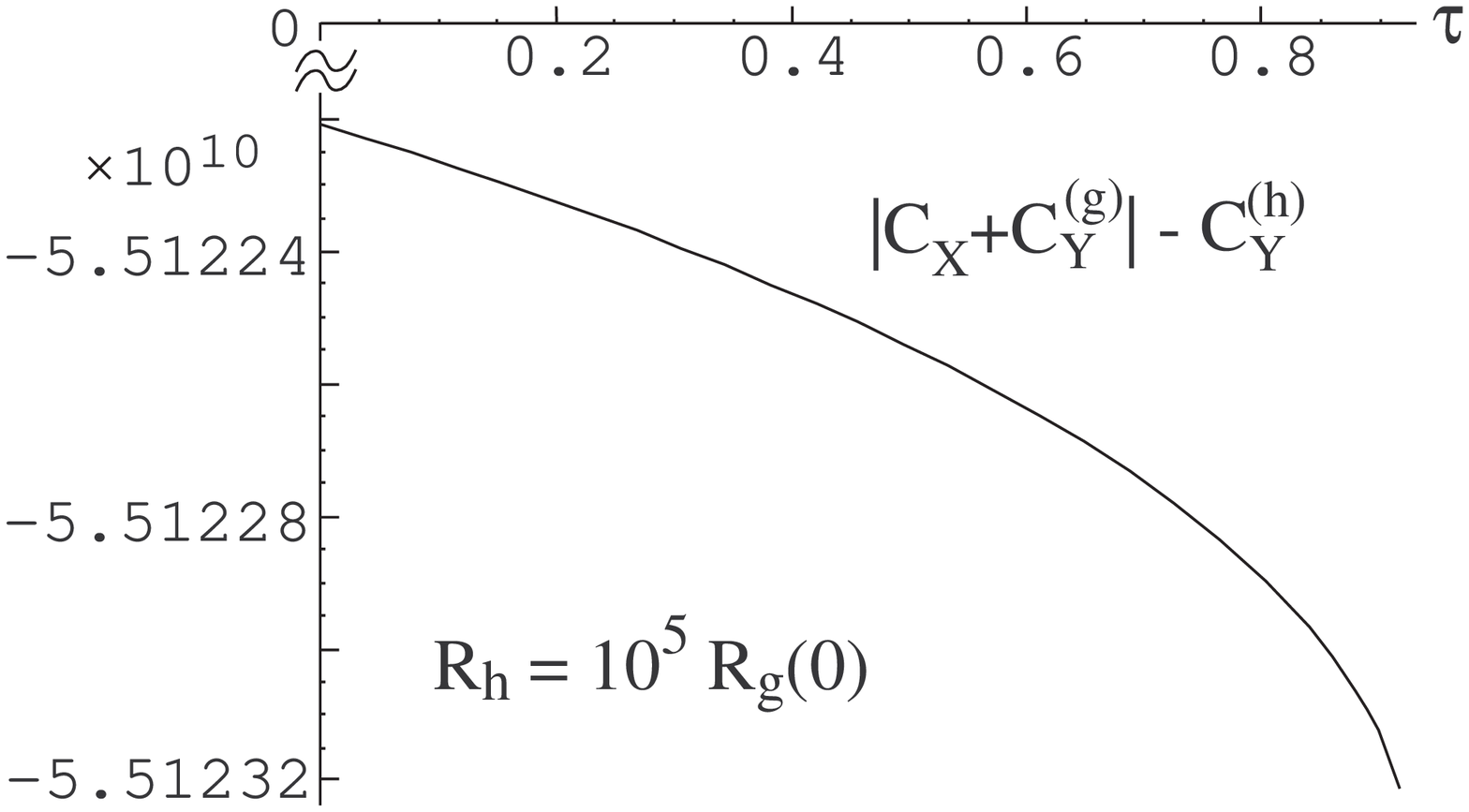}
 \end{center}
\caption{$| C_X + C_Y^{(g)} | - C_Y^{(h)}$ for the cases $R_h/R_g(0) = 100$, $10^3$ and $10^5$ in figure \ref{fig-2}.}
\label{fig-3}
\end{figure}

Figure \ref{fig-3} shows time evolutions of $| C_X + C_Y^{(g)} | - C_Y^{(h)}$ for the cases $R_h/R_g(0) = 100$, $10^3$ and $10^5$. It is found from this figure that, as discussed at the end of section \ref{sec-ne.energy}, inequality \eqref{eq-ne.criterion} is not always satisfied during a black hole evaporation process. Even if inequality \eqref{eq-ne.criterion} is not satisfied at initial time, it must come to be satisfied in the course of evaporation process. When inequality \eqref{eq-ne.criterion} is not satisfied at initial time, the time $\tn$ tends to be longer.

\begin{figure}[t]
 \begin{center}
  \includegraphics[height=30mm]{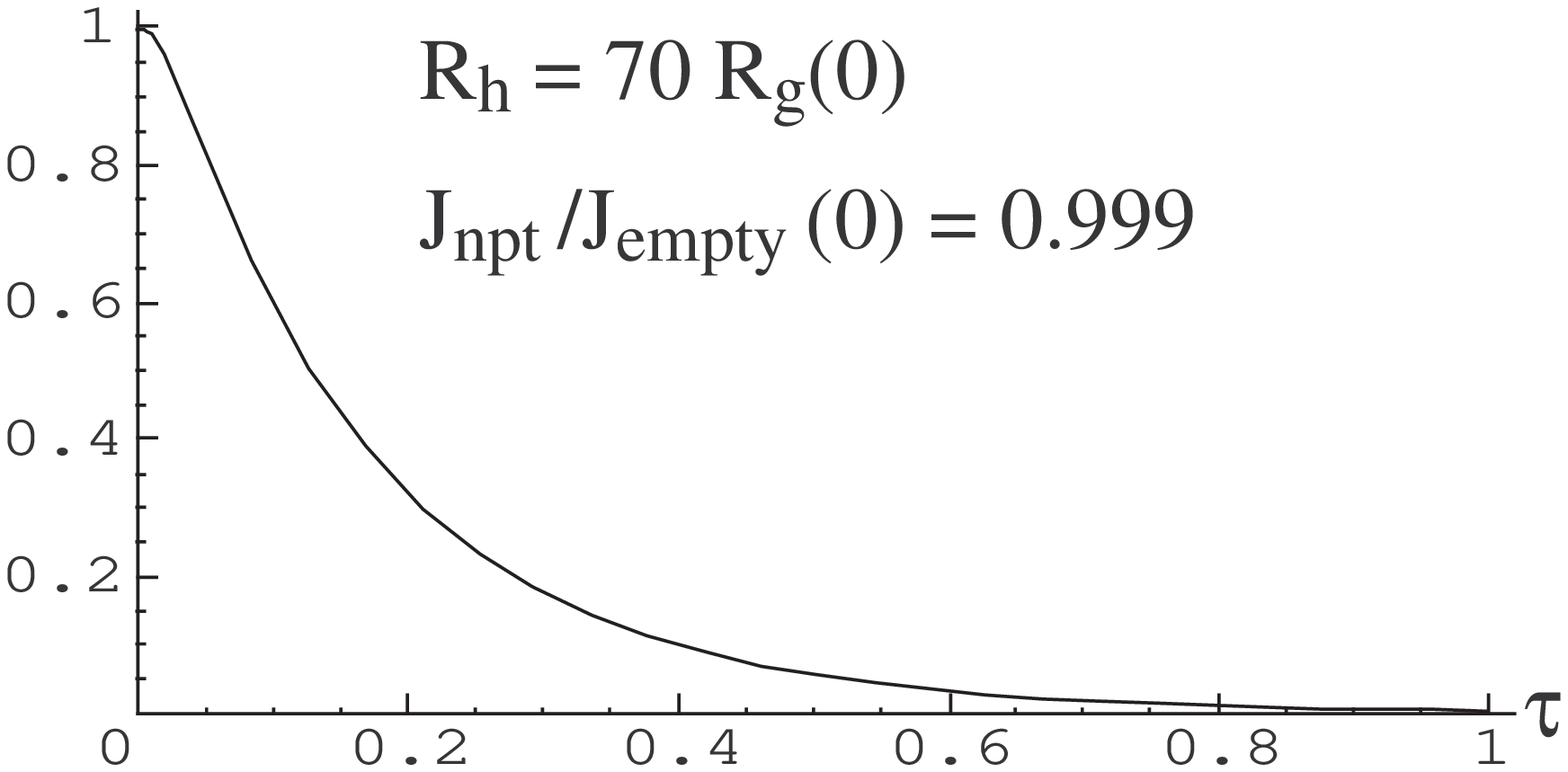}
  \includegraphics[height=30mm]{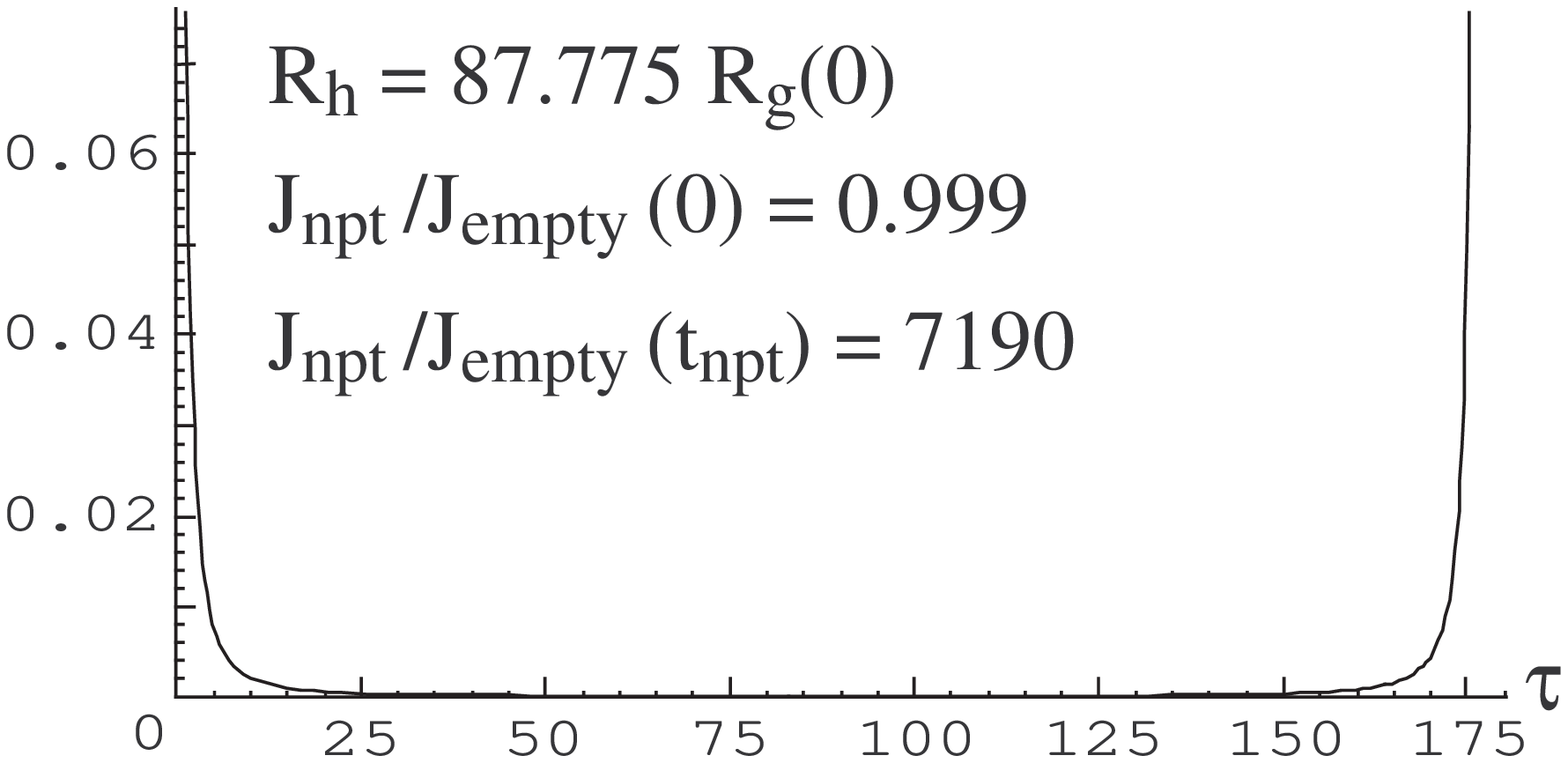}
  \includegraphics[height=30mm]{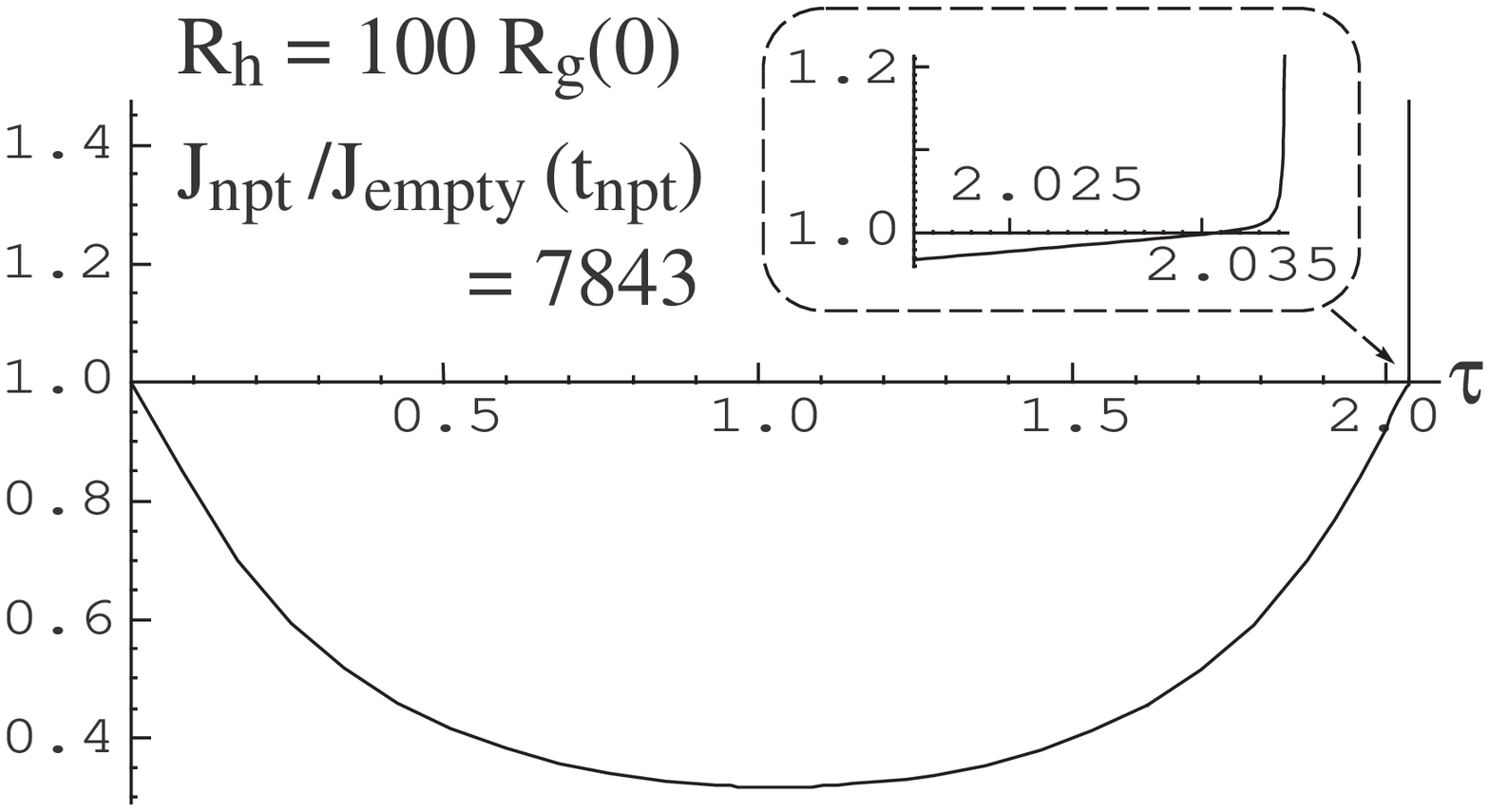} \\
  \includegraphics[height=30mm]{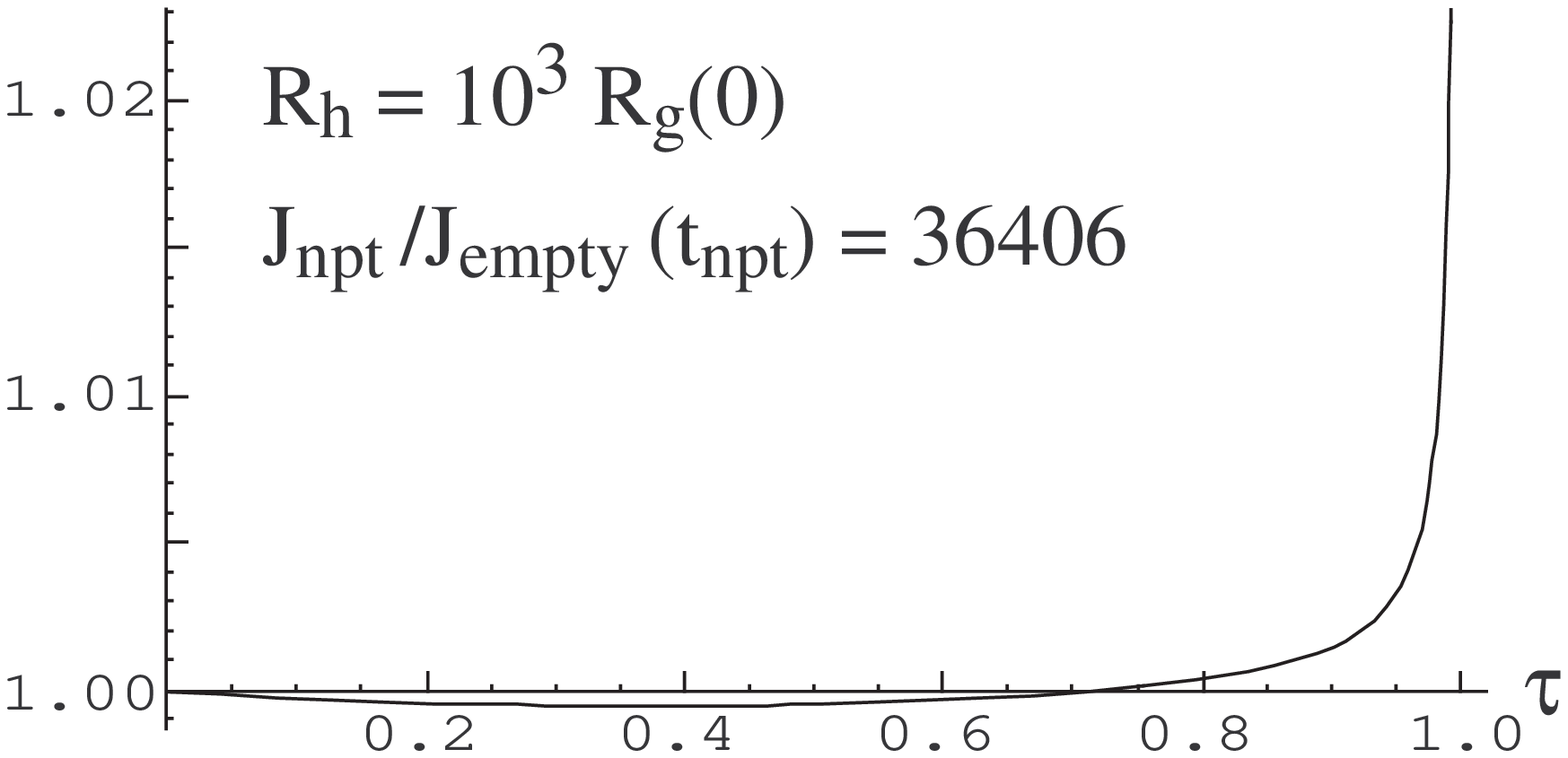}
  \includegraphics[height=30mm]{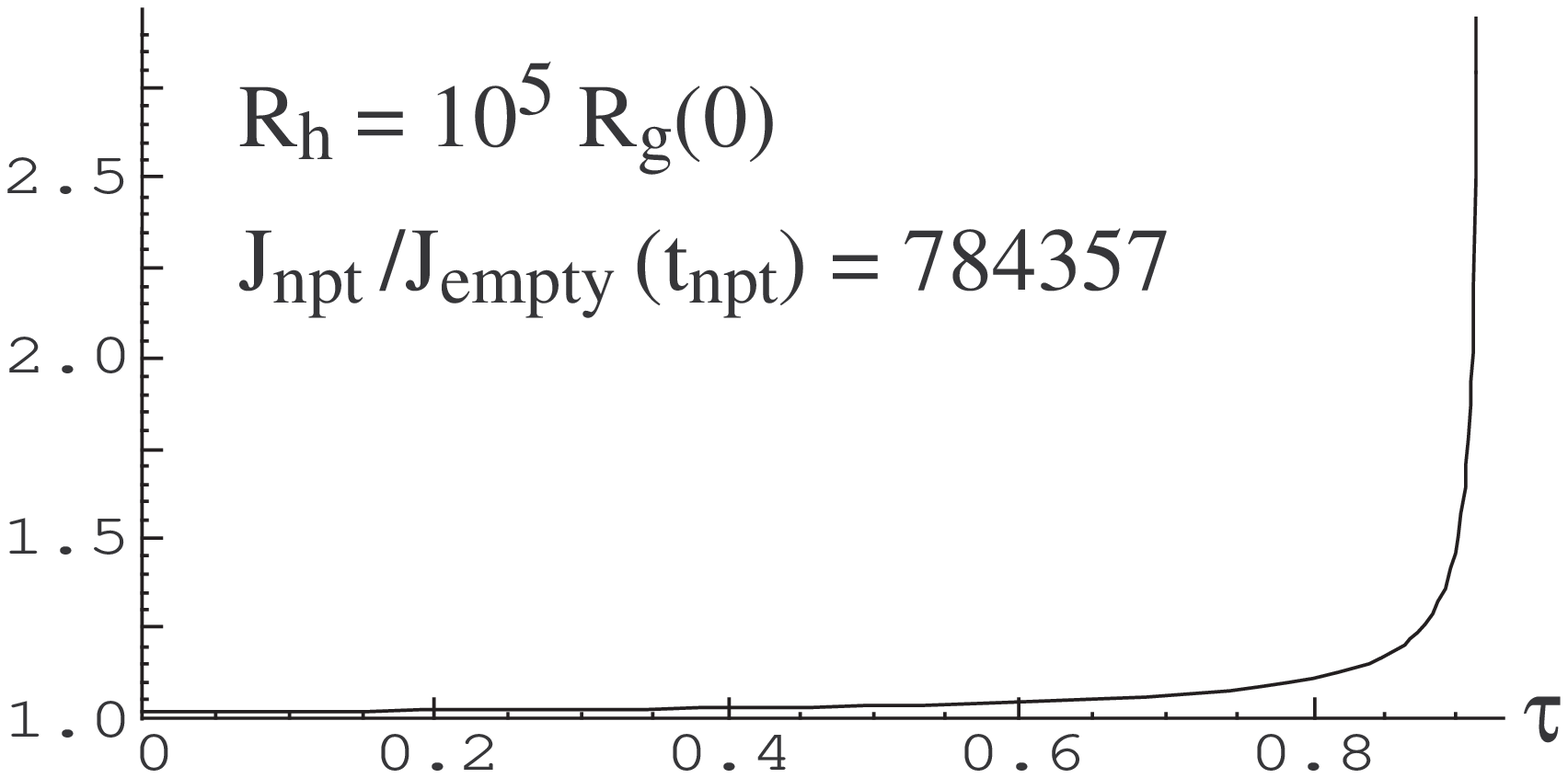}
  \includegraphics[height=30mm]{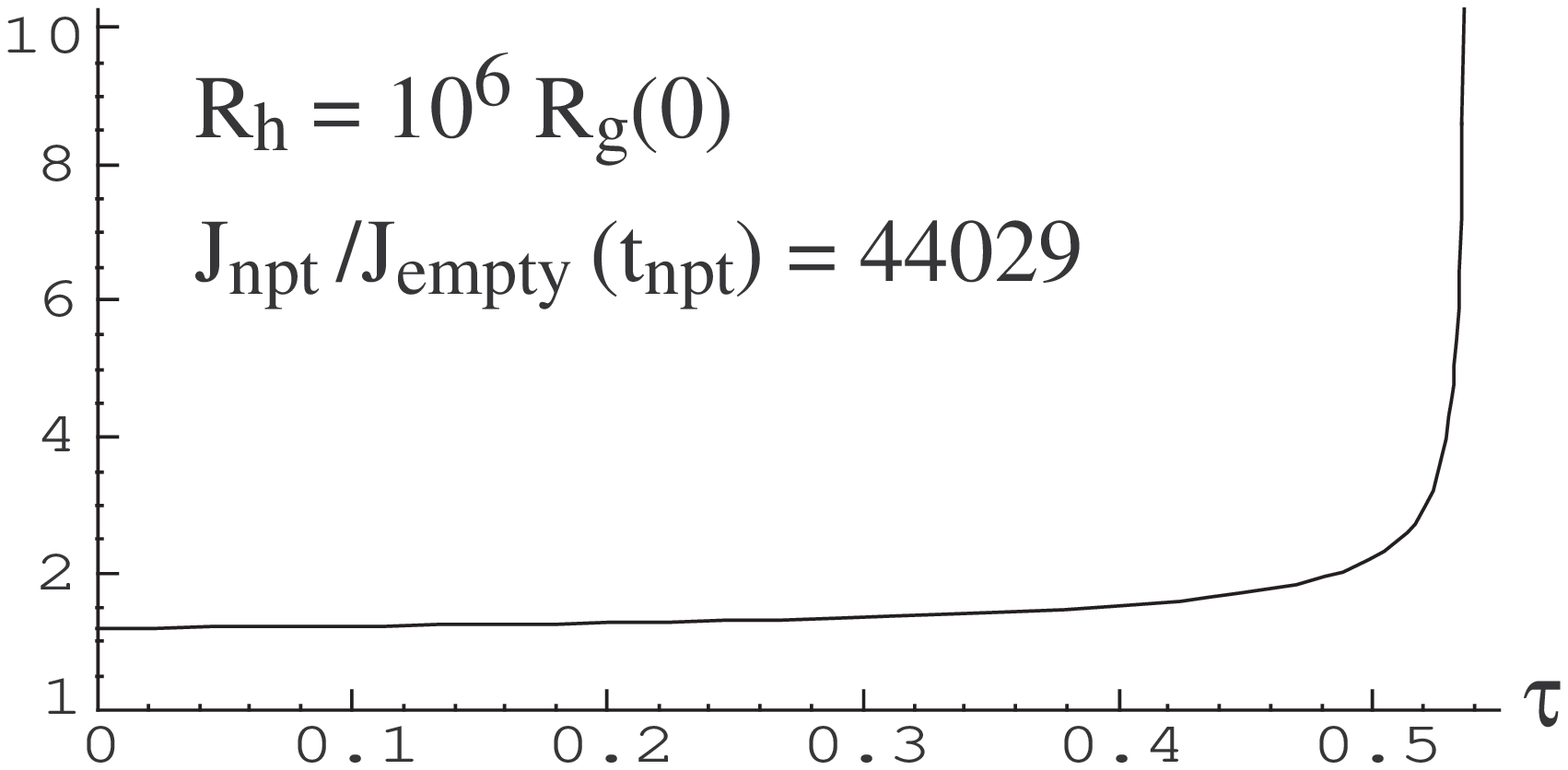}
 \end{center}
\caption{$\jn/\je$ for each case in figure \ref{fig-2}. The graph in upper-right panel is continuous and smooth.}
\label{fig-4}
\end{figure}

Figure \ref{fig-4} shows time evolutions of the ratio of energy emission rates, $\jn/\je = \left(C_g/C_X\right) \left( 1 - T_h^4/T_g^4 \right)$, given by equation \eqref{eq-ne.jn.je}. It is recognised from this figure that the expectation given at the end of previous subsection is supported. Figure \ref{fig-4} indicates that, when a black hole evaporates, the relation $\jn > \je$ comes to hold even if the converse relation $\jn < \je$ holds at initial time. For the case $R_h = 70 R_g(0)$ in which a black hole stabilises with a heat bath, the energy emission rate $\jn$ disappears, $\jn/\je \to 0$, as easily expected by the behaviour $T_g \to T_h \, \Rightarrow \, \jn \to 0$.

\subsubsection{Abrupt catastrophic evaporation}

In order to find a specific nonequilibrium phenomenon of NPT model, we point out a numerical evidence shown in list \eqref{eq-ne.list} that the black hole radius $R_g^{\ast}$ at time $\tn$ is greater than unity $R_g^{\ast} > 1$. According to the discussion in section \ref{sec-ne.energy} (the paragraph one before the last one), when a black hole evaporates in the framework of NPT model, a highly nonequilibrium dynamical stage of evaporation process occurs at semi-classical level $R_g^{\ast} > 1$. After the time $\tn$, a black hole and radiation fields should be described as highly dynamical ones.

In the following discussion, we make two steps: firstly, we show analytically that $R_g^{\ast} > O(1)$ holds with a rather general condition. Secondly, a physical implication of $R_g^{\ast} > O(1)$ is discussed. Then, a specific nonequilibrium phenomenon of NPT model is suggested.

For the first, we analyse the energy transport equations \eqref{eq-ne.transport.2}. Because of definition \eqref{eq-ne.tnpt} of time $\tn$, we consider two cases, $\tn = t_1$ and $\tn = t_2$, where $t_1$ and $t_2$ are given in definition \eqref{eq-ne.tnpt}. But before proceeding to the analysis of these cases, we should point out the following: as explained at the end of appendix \ref{app-CXCYg}, there is a possibility that equation $C_X(R_g) + C_Y^{(g)}(R_g) = 0$ has one, two or three solutions of $R_g$ for a sufficiently small $T_h$. However, as discussed at the end of appendix \ref{app-CXCYg}, even if there are two or three solutions of $R_g$ for our choice of $T_h$, it is the lowest solution at which the quasi-equilibrium assumption is violated. Hence we consider the time $t_1$ in definition \eqref{eq-ne.tnpt} of $\tn$ as the lowest solution of equation $C_X(t) + C_Y^{(g)}(t) = 0$.

Here we estimate the order of $R_g^{\ast}$. Consider the case $\tn = t_1$, where $R_g^{\ast} = R_g(t_1)$ and $C_X(R_g^{\ast}) + C_Y^{(g)}(R_g^{\ast}) = 0$. Because of $C_Y^{(g)} > 0$ by definition, $C_X(R_g^{\ast}) = - C_Y^{(g)}(R_g^{\ast}) < 0$ holds. Consequently, according to a behaviour of $C_X(R_g)$ explained in appendix \ref{app-CX.Rg} (left panel in figure \ref{fig-app.2}), we find $R_g^{\ast} > \tilde{R}_g \simeq 0.055 \times \left( N R_h \right)^{1/3}$. Hence together with equation \eqref{eq-ne.N}, we find $R_g^{\ast} \gtrsim O(1)$ for the situation $R_h \gtrsim O(10^3)$. And next consider the other case $\tn = t_2$, where $|\dot{R}_g(t_2)| = 1$. Because of $t_2 < t_1$, we find $C_X(t_2) + C_Y^{(g)}(t_2) < 0$. Then, because it is assumed that $T_h$ is small enough so that the validity condition \eqref{eq-ne.validity.1} holds, we find by appendix \ref{app-CXCYg} (figure \ref{fig-app.3}) that $R_g^{\ast} = R_g(t_2) > R_{g0}$ holds, where $R_{g0}$ is given by $C_X(R_{g0}) + C_Y^{(g)}(R_{g0}) = 0$. Therefore, following the same discussion given for the case $\tn = t_1$, we obtain $R_g^{\ast} > R_{g0} > \tilde{R}_g \simeq O(1) \, \Rightarrow \, R_g^{\ast} \gtrsim O(1)$ for the situation $R_h \gtrsim O(10^3)$. In summary, the black hole radius $R_g^{\ast}$ at time $\tn$ is greater than order unity $R_g^{\ast} > O(1)$, when the black hole evaporates in the framework of NPT model with the condition $R_h \gtrsim O(10^3)$. Here we have to note two points: first is that the condition $R_h \gtrsim O(10^3)$ is not a necessary condition but a sufficient condition for $R_g^{\ast} > O(1)$, and there remains a possibility that $R_g^{\ast} > O(1)$ holds even if $R_h < O(10^3)$. Second point is that, if the nonequilibrium nature of black hole evaporation is not taken into account, the radius $R_g^{\ast}$ can not be greater than order unity but it becomes less than Planck length (see appendix \ref{app-final}). The relation $R_g^{\ast} > O(1)$ is peculiar to the NPT model.

We proceed to the second part of this subsection, an implication of the above result $R_g^{\ast} > O(1)$. Recall that a highly nonequilibrium dynamical stage of evaporation process occurs after the time $\tn$. (In the following, a ``highly dynamical evaporation stage'' means a ``highly nonequilibrium dynamical stage of black hole evaporation process''.) Because of $R_g^{\ast} > O(1)$, a semi-classical (but not quasi-equilibrium) discussion can be available for the highly dynamical evaporation stage while the black hole radius shrinks from $R_g^{\ast}$ ($> O(1)$) to $R_p$ ($\sim 1$, Planck length). Then it is appropriate to consider that the mass energy of black hole evolves from $E_g^{\ast}$ ($= R_g^{\ast}/2$) to $E_p$ ($= R_p/2 \sim 1$, Planck energy), and that the speed of shrinkage of the horizon $v = \left|dR_g/dt\right|$ is approximately unity $v \sim 1$ during the highly dynamical evaporation stage. The energy $\Delta E_g = E_g^{\ast} - E_p$ is emitted during the highly dynamical evaporation stage. Further, for example, figure \ref{fig-4} of our numerical example in previous subsection implies a very strong luminosity $\jn$ of Hawking radiation in the NPT model in comparison with the luminosity $\je$ in the evaporation in an empty space. And note that the luminosity $\je$ in an empty space is very strong as explained in section 1 of reference \cite{ref-hr}. Hence, $\jn$ may be a huge luminosity. That is, the energy emission by a black hole in the NPT model may be understand as a strong ``burst''. In addition to the luminosity of Hawking radiation, we consider the duration $\delta t$ of the highly dynamical evaporation stage. The duration $\delta t$ is estimated as $\delta t \sim R_g^{\ast}/v \sim R_g^{\ast}$, and the following relation is obtained,
\sikib
 \delta\tau \equiv \frac{\delta t}{\te} \sim \frac{R_g^{\ast}}{\te}
 < \frac{R_g(0)}{\te} = \frac{N}{1280 \pi R_g(0)^2} \sim \frac{1}{400 R_g(0)^2} \ll O(0.01) \, ,
\label{eq-ne.duration}
\sikie
where we set $N \sim 10$ due to equation \eqref{eq-ne.N}, and used $R_g^{\ast} < R_g(0)$ and $1 \ll R_g(0)$ (the initial radius $R_g(0)$ should be large enough to consider a semi-classical evaporation). This denotes $\delta t \ll \te$. Further, for example, we find $\delta\tau \ll O(0.01) < \tn/\te$  ($\Rightarrow \, \delta t \ll \tn$) for each panel in figure \ref{fig-2}. Hence it seems reasonable to consider that $\delta t$ is very shorter than $\tn$. (For example it seems that the tangent $dT_g/dt$ seen in each panel in figure \ref{fig-2} becomes very large quickly as $t \to \tn$, and $T_g$ will reach Planck temperature quickly just after $\tn$.) Hence it is concluded that the energy $\Delta E_g = E_g^{\ast} - E_p$ bursts out of black hole with a very strong luminosity within $\delta t$ which is negligibly short in comparison with $\tn$. 

From the above, we can suggest the following picture: when a black hole evaporates in the framework of NPT model with a condition $R_h \gtrsim O(10^3)$, a quasi-equilibrium evaporation stage continues until $\tn$. Then a highly dynamical evaporation stage occurs at $\tn$. In the highly dynamical evaporation stage, a semi-classical black hole of radius $R_g^{\ast} > O(1)$ evaporates abruptly (within a negligibly short time scale $\delta t$) to become a quantum one. This abrupt evaporation in the highly dynamical evaporation stage is accompanied by a burst of energy $\Delta E_g$. We call this phenomenon {\it the abrupt catastrophic evaporation} at semi-classical level $R_g^{\ast} > O(1)$, where ``catastrophic'' means that the speed of shrinkage of horizon is very high $v \sim 1$ and the energy $\Delta E_g$ bursts out of black hole with a huge luminosity within a negligibly short time scale $\delta t$ in comparison with $\tn$.

The above discussion is based on the NPT model. As shown in appendix \ref{app-final}, in the equilibrium model used in section \ref{sec-eq} and in the model that a black hole evaporates in an empty space, the black hole radius becomes Planck size before the shrinkage speed of horizon reaches unity. The highly dynamical evaporation stage and the abrupt catastrophic evaporation at semi-classical level do not occur in both models. Hence the abrupt catastrophic evaporation at semi-classical level seems to be a specific nonequilibrium phenomenon of NPT model. 

Finally we estimate a typical time scale of black hole evaporation in a heat bath. It is reasonable to consider the duration of quantum evaporation stage is about Planck time. Then, a time scale of evaporation is given by $\tn + \delta t + O(1) \sim \tn$. The time $\tn$ gives a typical time scale of black hole evaporation in a heat bath.

\subsubsection{Conclusion of section \ref{sec-ne.evapo}}

From the above, we find the nonequilibrium effect of energy exchange on a black hole evaporation in a heat bath as follows: {\it When the nonequilibrium region around black hole is not so large, the evaporation time scale $\tn$ in a heat bath becomes longer than the evaporation time scale $\te$ in an empty space (ignoring gravitational interactions between black hole and radiation fields). However, as the nonequilibrium region around black hole is set larger, the time scale $\tn$ becomes shorter than the time scale $\te$. That is, the nonequilibrium effect of energy exchange tends to accelerate the black hole evaporation. Further, for the case $R_h \gtrsim O(10^3)$, a quasi-equilibrium evaporation process proceeds to a quantum evaporation stage abruptly at $\tn$ at semi-classical level $R_g^{\ast} > O(1)$ with a burst of energy (the abrupt catastrophic evaporation).} 

Finally in this section, we discuss about a black hole evaporation in an empty space in a full general relativistic framework. Note that, even if a black hole is in an empty space, there should exists an incoming energy flow into the black hole due to the curvature scattering. When the curvature scattering is taken into account for the case of a black hole evaporation in an empty space, we can interpret the whole system as that a black hole is surrounded by some nonequilibrium matter fields which possess outgoing and incoming energy flows of Hawking radiation under the effects of curvature scattering. Further, since the curvature scattering occurs whole over the spacetime, it is expected that the nonequilibrium region is so large that a condition which corresponds to $R_h \gtrsim O(10^3)$ in the NPT model holds. Hence, if the NPT model is extended to a full general relativistic model, we can expect that a black hole evaporation in an empty space can be treated in the framework of full general relativistic version of the NPT model (with removing the heat bath), and that the abrupt catastrophic evaporation at semi-classical level may occur as well since the nonequilibrium region is sufficiently large.

\subsection{Final fate of black hole evaporation}
\label{sec-ne.entropy}

In previous section, the black hole evaporation processes before a quantum evaporation stage have been considered. In this section, we consider whether some remnant remains after the quantum evaporation stage or not. The study on the final fate of black hole evaporation, which has a close relation with the information loss problem \cite{ref-info}, is usually carried out in the context of quantum gravity. However, since no complete theory of quantum gravity has yet been constructed, it is meaningful to some extent to study the final fate of black hole evaporation with an appropriate model of black hole evaporation without referring to present incomplete theories of quantum gravity. Therefore, we use the NPT model and try to extract what we can suggest about the final fate of black hole evaporation.

For the time being, we assume that a black hole evaporates out completely and only equilibrium radiation fields remain as the end state of quantum evaporation stage. If a contradiction results from this assumption, we may conclude that a remnant will remain after a black hole evaporation. This section aims to suggest a necessity of some remnant by the reductive absurdity.

As discussed at the end of section \ref{sec-ne.evapo}, it can be expected that the abrupt catastrophic evaporation at semi-classical level occurs in a full general relativistic framework. Therefore we consider the case $R_h \gtrsim O(10^3)$ in the NPT model which denotes the occurrence of abrupt catastrophic evaporation at semi-classical level. Then, under the assumption of complete evaporation of black hole, we can draw a scenario in the framework of NPT model as follows. 

\begin{figure}[t]
 \begin{center}
 \includegraphics[height=35mm]{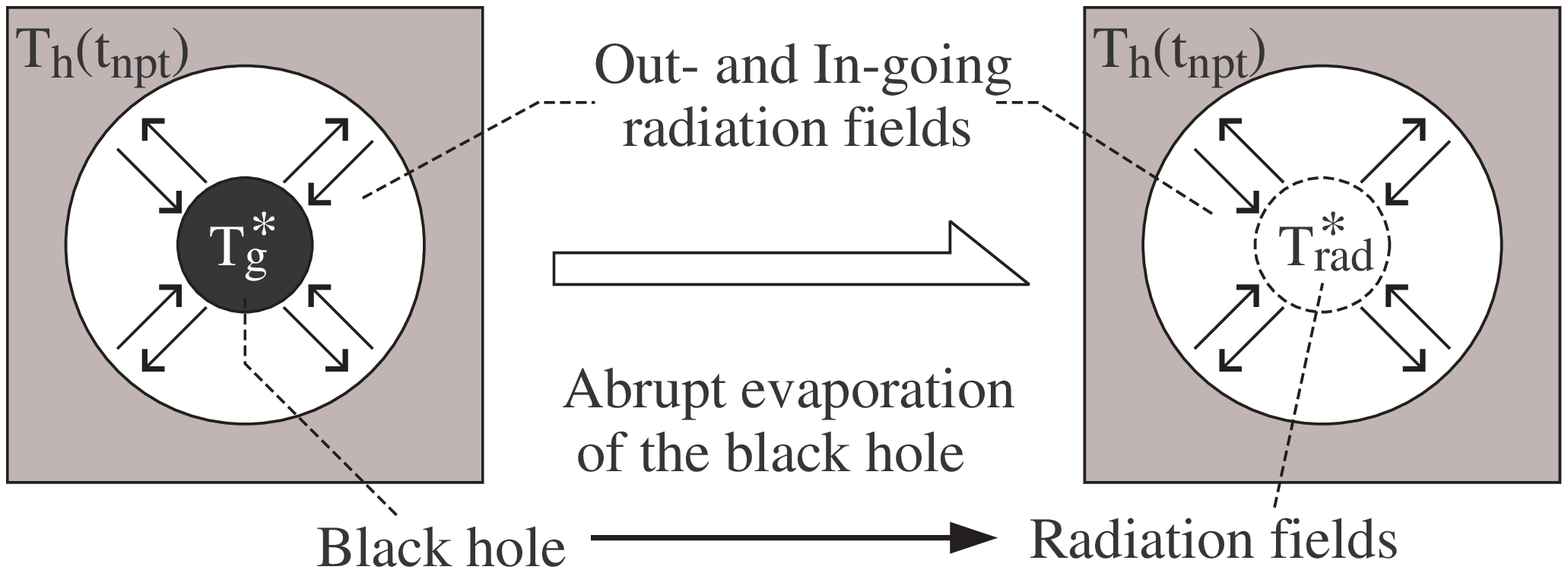} \qquad
 \includegraphics[height=30mm]{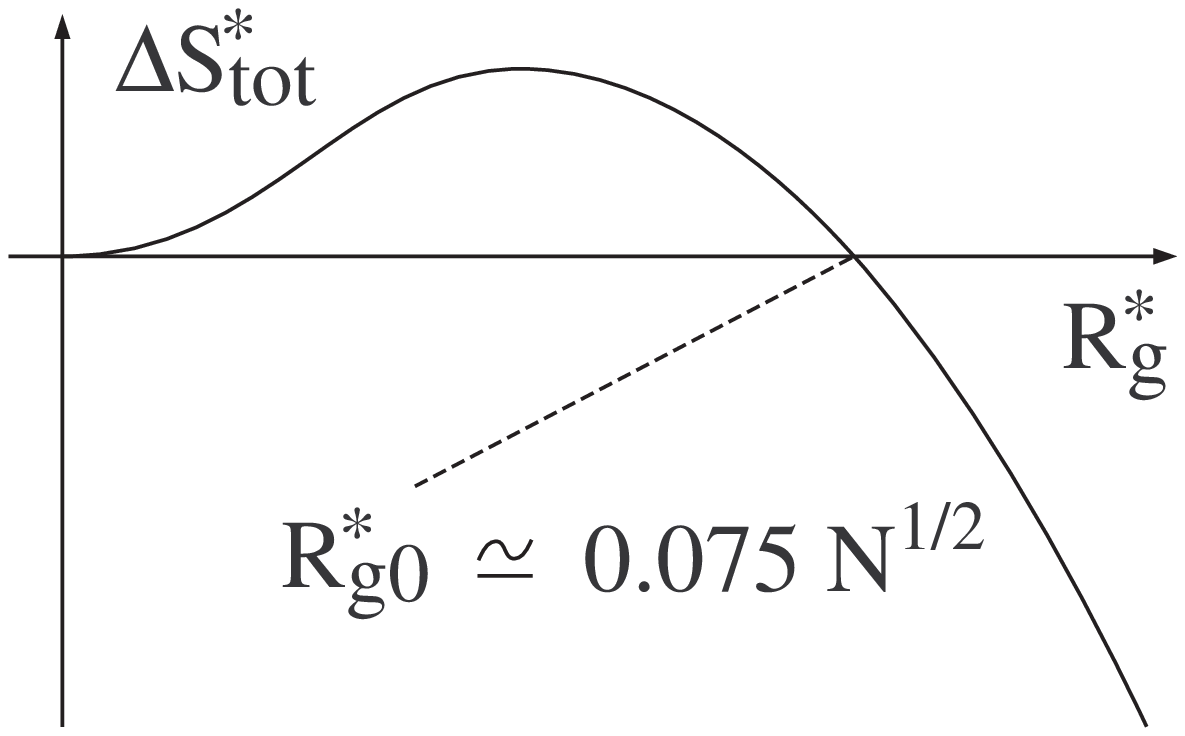}
 \end{center}
\caption{If a black hole evaporates out completely, when a black hole temperature $T_g$ reaches $T_g^{\ast}$, the black hole is suddenly replaced by radiation fields of temperature $T_{rad}^{\ast}$ which is determined by the energy conservation. However it is impossible from entropic viewpoint.}
\label{fig-5}
\end{figure}

As a quasi-equilibrium stage of black hole evaporation proceeds until $\tn$, the temperature $T_g$ approaches a critical value $T_g^{\ast} = 1/4 \pi R_g^{\ast}$, where $R_g^{\ast} = R_g(\tn)$. Then, a highly dynamical evaporation stage and a quantum evaporation stage follow. Under the assumption of complete evaporation, these successive stages are described as one phenomenon; a black hole emits completely its mass energy $E_g^{\ast}$ ($= R_g^{\ast}/2$) within a negligibly short time scale just after the time $\tn$ (abrupt catastrophic evaporation), and equilibrium radiation fields remain. In the NPT model, this abrupt catastrophic evaporation under the assumption of complete evaporation is treated by the procedure; replace a black hole of radius $R_g^{\ast}$ with equilibrium radiation fields of volume $V_g^{\ast} = (4 \pi/3) \, R_g^{\ast \, 3}$. In this replacement, thermodynamic states of heat bath and nonequilibrium radiation fields around the region of $V_g^{\ast}$ are not changed (see left panel in figure \ref{fig-5}). Here note that the energy of equilibrium radiation fields in volume $V_g^{\ast}$ should equal the mass energy $E_g^{\ast}$ of black hole. Therefore the temperature $T_{rad}^{\ast}$ of equilibrium radiation fields is determined by
\sikib
 E_g^{\ast} = 4 \, \sigmap \, T_{rad}^{\ast \, 4} \, V_g^{\ast} \quad \Rightarrow \quad
 T_{rad}^{\ast} = \left(\, \frac{3}{32 \, \sigmap \, \pi \, R_g^{\ast \, 2}} \,\right)^{1/4} \, ,
\label{eq-ne.temperature.rad}
\sikie
where $4 \sigmap \, T^4 \, V$ is the equilibrium energy of radiation fields (see appendix \ref{app-sst}). When a sufficiently long time has passed after the abrupt catastrophic evaporation (under the assumption of complete evaporation), the whole system reaches an equilibrium state in which a heat bath and radiation fields in the hollow region have the same equilibrium temperature. However, without considering this totally equilibrium end state of the whole system, but with considering the states just before and just after the abrupt catastrophic evaporation, we can discuss whether a remnant remains or not after a black hole evaporation.

Here we should notice that the total entropy must increase along a black hole evaporation process, since the whole system which consists of a black hole, a heat bath and radiation fields is isolated from outside environments. Therefore the relation $S_{tot}^{\ast} < S_{tot}^{\ast \, \prime}$ should hold, where $S_{tot}^{\ast}$ is the total entropy of the whole system at time $\tn$ just before the abrupt catastrophic evaporation, and $S_{tot}^{\ast \, \prime}$ is the total entropy of the whole system just after the complete evaporation of black hole. Because the abrupt catastrophic evaporation is described by a simple replacement of a black hole with equilibrium radiation fields, we find that the entropy difference $\Delta S_{tot}^{\ast} \equiv S_{tot}^{\ast \, \prime} - S_{tot}^{\ast}$ is given by
\sikib
 \Delta S_{tot}^{\ast} \equiv S_{tot}^{\ast \, \prime} - S_{tot}^{\ast}
 = S_{rad}^{\ast} - S_g^{\ast} \, ,
\sikie
where $S_g^{\ast}$ is the black hole entropy of temperature $T_g^{\ast}$ and $S_{rad}^{\ast}$ is the equilibrium entropy of radiation fields of volume $V_g^{\ast}$ and temperature $T_{rad}^{\ast}$. If the complete evaporation of black hole is true of the case, a relation $\Delta S_{tot}^{\ast} > 0$ must hold. However it is impossible as shown below.

Using equations of states \eqref{eq-eq.eos} and the equilibrium entropy of radiation fields $(16 \sigmap/3)\, T^3 \, V$ together with the temperature $T_{rad}^{\ast}$ of equation \eqref{eq-ne.temperature.rad}, the entropy difference is calculated as
\sikib
 \Delta S_{tot}^{\ast}
 = \frac{16 \, \sigmap}{3} \, T_{rad}^{\ast \, 3} \, V_g^{\ast} - S_g^{\ast}
 = \frac{2}{3} \left(\, \frac{32 \, \pi \, \sigmap}{3} \,\right)^{1/4} R_g^{\ast \, 3/2}
   - \pi \, R_g^{\ast \, 2} \, .
\label{eq-ne.entropy.diff}
\sikie
A schematic graph of $\Delta S_{tot}^{\ast}$ is shown in right panel in figure \ref{fig-5}, where $R^{\ast}_{g 0} = (8/27 \sqrt{5 \pi}) \sqrt{N} \simeq 0.075 \times \sqrt{N}$ is given by $\Delta S_{tot}^{\ast} = 0$ and $\sigmap = N\, \pi^2/120$. We find $\Delta S_{tot}^{\ast} > 0$ for $R_g^{\ast} < R_{g 0}^{\ast}$ and $\Delta S_{tot}^{\ast} < 0$ for $R_g^{\ast} > R_{g 0}^{\ast}$. Here recall that, because of $C_Y^{(g)} > 0$ by definition and a behaviour of $C_X(R_g)$ explained in appendix \ref{app-CX.Rg} (left panel in figure \ref{fig-app.2}), an inequality $R_g^{\ast} > \tilde{R}_g$ holds due to the validity condition \eqref{eq-ne.validity.1}, where $\tilde{R}_g$ is given by solving $C_X = 0$ for $R_g$. Further, using equation \eqref{eq-CX.tRg}, we find $\tilde{R}_g/R_{g 0}^{\ast} \simeq (27 \sqrt{5 \pi}/32 (30 \pi)^{1/3} ) \times (R_h^2/N)^{1/6} \simeq 0.734 \times (R_h^2/N)^{1/6}$. Then, by equation \eqref{eq-ne.N} and requirement $R_h \gtrsim O(10^3)$ of the occurrence of abrupt catastrophic evaporation, we obtain
\sikib
 \frac{R_g^{\ast}}{R_{g 0}^{\ast}} > \frac{\tilde{R}_g}{R_{g 0}^{\ast}} = O(1) \, .
\label{eq-ne.entropy.absurd}
\sikie
This indicates $\Delta S_{tot}^{\ast} < 0$. Consequently, by the reductive absurdity as explained at the beginning of this section, a complete evaporation of black hole is impossible.

From the above, we give a suggestion as follows: {\it a complete evaporation of black hole is prohibited and a remnant should remain after a black hole evaporation. The entropy of remnant should guarantee the increase of total entropy}.

We comment about the mass of remnant. One may think that the mass energy (or internal energy) of remnant may be extracted by applying an energetic analysis to the ``entropic'' discussion given above. One may obtain a minimum entropy $S_{min}^{(rem)}$ of remnant as $S_{min}^{(rem)} = S_g^{\ast}$, and a mass energy of remnant may be obtained from $S_{min}^{(rem)}$. However, because equations of states of remnant is unknown, it is impossible to obtain an energy of remnant from the entropic discussion. On the other hand, as discussed in section \ref{sec-ne.evapo} (before equation \eqref{eq-ne.duration}), the mass of quantum black hole is expected to be about one Planck energy. Therefore, as long as the NPT model is extrapolated over its validity, the mass of remnant seems to be of the order of one Planck energy.

In the rest of this section, we discuss more about the validity of discussion given in deriving inequality \eqref{eq-ne.entropy.absurd}. There are four points we are going to discuss here. For the first, we discuss about the entropic viewpoint in NPT model. As explained in appendix \ref{app-essence}, the NPT model is constructed referring to the validity of energetics of equilibrium model used in section \ref{sec-eq} and reference \cite{ref-trans.1}. Therefore one may think that the NPT model can not reflect the entropy of gravitational field around a black hole. However, as mentioned in equation \eqref{eq-eq.capacity}, using the Euclidean path integral method for a black hole spacetime \cite{ref-entropy}, the entropy of gravitational field itself is just given by the quantity $S_g$ appeared in equations \eqref{eq-eq.eos}. That is, the black hole entropy is the entropy of the whole gravitational field in a black hole space time. Hence it seems that an entropic discussion in NPT model is appropriate.

Next for the second point, we discuss about the usage of NPT model in this section. If we want to analyse details of the final stage of black hole evaporation process, then a quantum gravity is necessary, since a black hole becomes quantum, $R_g < 1$. On the other hand, because of the quasi-equilibrium assumption, the black hole radius has to be restricted as $( R_g(0) > ) R_g^{\ast} > O(1)$ in the framework of NPT model. However this restriction $R_g^{\ast} > O(1)$ does not mean that the NPT model is never available for considering the final fate of black hole evaporation. Although the NPT model is based on a classical Schwarzschild black hole, once one assume that a black hole will evaporate out completely, then it is true under this assumption that radiation fields remain and its entropy can be counted after a quantum evaporation process ended. That is, even though we never refer to any details of present incomplete theories of quantum gravity, we can compare an entropy of black hole before the start of quantum evaporation process with an entropy of radiation fields after the end of quantum evaporation process. Hence our analysis done above seems to be reasonable to approach the final fate of black hole evaporation.

For the third point, we turn our discussion to the starting assumption that a black hole evaporates out completely and equilibrium radiation fields remain. As seen above, equation \eqref{eq-ne.entropy.diff} has led a contradiction to deny the assumption and to result in necessity of a remnant. Here recall that equation \eqref{eq-ne.entropy.diff} is calculated with using the equilibrium entropy of radiation fields. Then, one may think that it is more general to modify the starting assumption so that radiation fields after the complete evaporation of black hole are not necessarily in an equilibrium state. However, as explained in detail in reference \cite{ref-sst}, the nonequilibrium entropy of radiation fields should be smaller than the equilibrium one $S_{rad}^{\ast}\, ( < S_g^{\ast} )$ used above, since the equilibrium state is the maximum entropy state. That is, it is reasonable to expect that an inequality $\Delta S_{tot}^{\ast} < 0$ holds stronger for a modified starting assumption which requires nonequilibrium radiation fields after a complete evaporation of black hole.

Finally we discuss about a black hole evaporation in an empty space (a situation without heat bath). As discussed at the end of section \ref{sec-ne.evapo}, the abrupt catastrophic evaporation at semi-classical level is expected to occur for a black hole evaporation in an empty space in a full general relativistic framework. Further note that calculations of entropy difference $\Delta S_{tot}^{\ast}$ in equation \eqref{eq-ne.entropy.diff} do not depend on the outside of black hole, but depend only on the black hole entropy $S_g^{\ast}$ and the equilibrium entropy $S_{rad}^{\ast}$ of radiation fields. That is, without respect to the degree of nonequilibrium nature of the environment around black hole, once the abrupt catastrophic evaporation of black hole occurs, the same discussion as given in this section seems to be applicable to a black hole evaporation in an empty space. Therefore, a full general relativistic treatment of black hole evaporation in an empty space may result in necessity of a remnant at the end state of black hole evaporation.

\section{Summary and discussions}
\label{sec-sd}

When a black hole evaporates in a heat bath, there exists an energy flow between the black hole and the heat bath due to the Hawking radiation and the accretion of matters from heat bath to black hole. This denotes that the whole system which consists of a black hole and a heat bath is not in an equilibrium. We have investigated nonequilibrium effects of energy flow on the black hole phase transition. Our analysis is carried out with the NPT model which reflects the nonequilibrium nature of energy exchange between a black hole and a heat bath. Then it is found that, contrary to the original equilibrium model used in reference \cite{ref-trans.1}, inequality \eqref{eq-ne.criterion} evaluated at initial time can not work as a criterion of the occurrence of instability of equilibrium state of the whole system. The nonequilibrium effect of energy exchange tends to make the situation more dynamical and complicated.

For the case that a black hole evaporates in a heat bath, the energetics of NPT model gives an understanding of nonequilibrium nature of black hole evaporation as follows:
\begin{itemize}
\item When the nonequilibrium region around black hole is not so large, the evaporation time scale $\tn$ in a heat bath becomes longer than the evaporation time scale $\te$ in an empty space, because of the incoming energy flow from the heat bath to the black hole. However, as the nonequilibrium region around black hole is set larger, the time scale $\tn$ becomes shorter than the time scale $\te$, because a nonequilibrium effect of temperature difference between the black hole and the heat bath appears as a strong energy extraction from the black hole by the heat bath. That is, the nonequilibrium effect of energy exchange between the black hole and the heat bath tends to accelerate the black hole evaporation. Further for the case $R_h \gtrsim O(10^3)$, the abrupt catastrophic evaporation at semi-classical level occurs.
\end{itemize}
Concerning a black hole evaporation in an empty space in a full general relativistic framework, we can expect this understanding holds due to the curvature scattering, as discussed at the end of section \ref{sec-ne.evapo}.

Two possible future applications of our discussion are suggested here. First possible future application is primordial black holes formed in radiation fields in the early universe \cite{ref-pbh}. The lower bound of the mass of primordial black holes which survive at present is estimated usually using the time scale $\te$ of black hole evaporation in an empty space. However, since the evaporation time scale $\tn$ in a heat bath changes according to how the radiation field in the early universe is far from an equilibrium, the lower bound of mass of primordial black holes may not be determined simply. Second possible future application is black holes created in particle accelerators in the context of TeV gravity \cite{ref-tev}. The black holes in particle accelerators, if they exist, will be created in a quark-gluon plasma in our brane and in a graviton gas in the higher dimensional spacetime. It is usually argued in many papers that, although a black hole in accelerator is highly dynamical at the moment of its creation, it will settle down to a stationary black hole before it proceeds to a quantum evaporation stage, since the evaporation time scale $\te$ in an empty space is slightly longer than the quasi-normal ringing time scale. However, since the evaporation time scale $\tn$ in a heat bath changes according to how the quark-gluon plasma and the graviton gas are far from an equilibrium at the moment of particle collision, the black holes in particle accelerators may not settle down to a stationary black hole before its quantum evaporation stage.

If a black hole is in an expanding universe as primordial black holes, not only the energy exchange between a black hole and its environments but also the cosmological expansion may cause nonequilibrium effects on black hole evaporation. Then one may think that the evaporation process may become a highly nonequilibrium one and be accelerated faster than the case without cosmological expansion. It is true that cosmological expansion disturbs the thermal nature of Hawking radiation and the spectrum of energy flow from black hole can not be treated as a Planckian distribution \cite{ref-bh.univ.1}. However nonequilibrium effects of cosmological expansion should require different considerations from the idea of this paper. Time evolution of the acceleration of cosmological expansion may be essential to determine whether the energy emission rate by black hole is enhanced or weakened. If the time derivative of acceleration of cosmological expansion is positive (or negative), the emission rate may be weakened (or enhanced) \cite{ref-bh.univ.2}. That is, when one consider details of black hole evaporation, the origin of nonequilibrium nature may be important. It seems that there are a variety of nonequilibrium effects. It is the nonequilibrium effect of energy exchange (or temperature difference) between a black hole and its environments that we investigates in this paper.

Next, concerning the end state of black hole evaporation, the entropic discussion of NPT model gives the suggestion as follow:
\begin{itemize}
\item Without respect to the size of nonequilibrium region around black hole, a remnant should remain after a black hole evaporation. The entropy of remnant should guarantee the increase of total entropy.
\end{itemize}
As long as the NPT model is extrapolated to a highly nonequilibrium dynamical stage and a quantum stage of evaporation after the time $\tn$, the mass of remnant seems to be of the order of one Planck energy. And, concerning a black hole evaporation in an empty space in a full general relativistic framework, we can expect the suggestion of necessity of remnant holds due to the curvature scattering, as discussed at the end of section \ref{sec-ne.entropy}. Further, recalling the third paragraph of section \ref{sec-intro}, our suggestion implies that a remnant can preserve the initial information of black hole formation, and that the unitary invariance of quantum theory is not violated. That is, the information loss problem \cite{ref-info} may not exist.

Finally we have to comment on another possibility (``non-remnant'' possibility) to preserve the unitary invariance along a black hole evaporation. The detailed properties of Hawking radiation may preserve the initial condition of gravitational collapse. Because it is the observation at an infinite future that a thermal radiation is observed in the Hawking radiation \cite{ref-hr}, a deviation of observed spectrum from Planckian distribution (at finite future from the onset of gravitational collapse) may carry the information of initial condition into somewhere on the spacetime, and the unitarity may be retained over the whole spacetime. However, if a naive and simple extrapolation of the NPT model to a quantum evaporation stage is valid (as assumed in section \ref{sec-ne.entropy}), then the thermodynamic state of outside environment of black hole is composed of a steady state of radiation fields and an equilibrium state of heat bath. Hence, because the steady state is described by a simple superposition of equilibrium states of different temperatures (see appendix \ref{app-sst}), it seems that a complicated (highly nonequilibrium) information of initial condition can not be preserved in steady states of radiation fields and equilibrium states of heat bath. Therefore, as far as the NPT model is extrapolated over its validity, a possible conclusion of $\Delta S_{tot}^{\ast} < 0$ (see equation \eqref{eq-ne.entropy.absurd}) is not a complicated nonequilibrium state of outside environment but a remnant.

\appendix
\section{On gravitational interactions between a black hole and a matter field}
\label{app-essence}

This appendix tries to give an intuitive explanation why a primitive model considered in section \ref{sec-eq} and reference \cite{ref-trans.1} retains the essence of black hole phase transition. Since section \ref{sec-eq} and reference \cite{ref-trans.1} do not include any relativistic effects except for equations of states \eqref{eq-eq.eos}, it may be expected that the essence of black hole phase transition is roughly explained with a Newtonian discussion.

But before proceeding to a Newtonian discussion, we consider the Hawking radiation by a Schwarzschild black hole in the context of general relativity. When a monochromatic wave mode of matter field of Hawking radiation propagates from a point near the black hole horizon to asymptotically flat region, a gravitational Doppler effect causes a red shift on the mode, $\omega_r = \omega/\sqrt{1-r_g/r}$, where $\omega_r$ and $\omega$ are respectively frequencies at starting point and at asymptotically flat region, $r_g$ is the gravitational radius of black hole, and $r$ is the radial coordinate (areal radius) at starting point. It is obvious $\omega_r \to \infty$ as $r \to r_g$. It has already been known that \cite{ref-tolman}, when a black hole is in a heat bath and they are in an equilibrium, the local equilibrium temperature $T(r)$ of heat bath at a point of radial coordinate $r$ should be given by $T(r) = T_g/\sqrt{1-r_g/r}$, where $T_g$ is a usual Hawking temperature of black hole which appears in equations \eqref{eq-eq.eos}, and the factor $\sqrt{1-r_g/r}$ is due to the gravitational red shift. We can understand $T(r)$ as that, in order to retain a thermal equilibrium against attractive gravitational force, a temperature which is higher than the asymptotic value should be required, since the higher temperature denotes the higher pressure against gravitational force.

Next we shift to a Newtonian discussion about matter field which composes a heat bath around black hole. Consider the case that the matter field is in an equilibrium with black hole. We put three assumptions: first is that the whole system which consists of the black hole and the matter field is isolated (total energy is conserved as the primitive model used in \cite{ref-trans.1}). Second assumption is that, for simplicity, the matter field of heat bath is the same with that of Hawking radiation. Third one is that self-interaction among particles of the matter field is negligibly weak in comparison with particle's kinetic energy and gravitational interaction with black hole.

Because of the assumption of very weak self-interaction, the energy $\varepsilon$ of one particle of the matter field is
\sikib
 \varepsilon = K(r) + U_g(r) \, , 
\sikie
where $K(r)$ is kinetic energy at radial distance $r$ from the center of black hole, and $U_g(r)$ is gravitational potential given by black hole. According to the gravitational Doppler effect $\omega_r = \omega/\sqrt{1-r_g/r}$, the gravitational potential should have the properties, $U_g \to - \infty$ as $r \to r_g$, and $U_g \to 0$ as $r \to \infty$. Then the total energy of the whole system is given by $E_{tot} \equiv E_g + \Sigma \varepsilon$, where $E_g$ is black hole mass, and a summation $\Sigma$ is taken for all composite particles of matter field. Here note that, since the whole system is in an equilibrium, a black hole is static and its mass is constant, $E_g = constatnt$. Further, because of the assumption that the whole system is isolated, we have $E_{tot} = constant$. Therefore we find $\Sigma \varepsilon = constant$, and together with the assumption of very weak interaction (energy is not exchanged among the particles), it is implied $\varepsilon = constant$. Then, due to the asymptotic value of $U_g(r\to\infty) = 0$, we find $\varepsilon = K(r) + U_g(r) = K_{\infty}$, where $K_{\infty} = K(r\to\infty)$ is the value of kinetic energy at infinity.

In order to understand the meaning of $K_{\infty}$, we consider the local temperature of matter field around black hole. Due to the law of equipartition of energy, we have
\sikib
 \left<\, K(r) \,\right> \propto T(r) \, ,
\label{eq-doppler.1}
\sikie
where $\left< K(r) \right>$ is a statistical mean value of $K(r)$ at $r$, and $T(r)$ is the local equilibrium temperature of matter field at $r$. Then, because of relation $K(r) = \varepsilon - U_g(r) = K_{\infty} - U_g(r)$, we find properties of local equilibrium temperature, $T(r) \to \infty$ as $r \to r_g$ and $T(r) \to T_g$ as $r \to \infty$. These behaviours of $T(r)$ correspond to those of a general relativistic relation $T(r) = T_g/\sqrt{1 - r_g/r}$. This may be interpreted as follows: when we consider separately a density of internal energy $e(r)$ of matter field (where $e(r) = n(r)\,\left< K(r) \right>$, $n(r)$ is the number density of particle at $r$) and a density of gravitational interaction energy $u(r)$ by black hole (where $u(r) = n(r)\,U_g(r)$), then infinite values appear $e(r\to r_g) = \infty$ and $u(r\to r_g) = -\infty$, and one may think thermodynamic considerations break down at $r = r_g$ due to the infinite values. However, the relation $K(r) + U_g(r) = K_{\infty}$ implies that these infinite values cancel each other. That is, although a realistic situation of matter field around black hole should be understood that its energy is influenced by the gravitational potential by black hole; however, when one considers an ``effective'' internal energy of matter field $E_{\infty} = N \left< K_{\infty} \right>$ (where $N$ is the number of particles of the matter field), then this energy $E_{\infty}$ equals the sum of the whole internal energy of the matter field and the interaction between black hole and particles of the matter field, $\Sigma\,\varepsilon = E_{\infty}$. That is, the infinite values of energies at $r = r_g$ are ``renormalised'' into the effective value $E_{\infty}$. This ``renormalised energy'' $E_{\infty}$ is the internal energy of the matter field evaluated by a thermal equilibrium state of asymptotic temperature $T_g$ without the influence of gravitational field of black hole. Hence, the total energy is calculated as $E_{tot} = E_g + \Sigma\,\varepsilon = E_g + E_{\infty}$.

From the above it is recognised that, as long as we use the renormalised internal energy $E_{\infty}$ of matter field of asymptotic temperature $T_g$, energetics of the whole system which consists of a black hole and matter field can be considered with ignoring the gravitational interaction between black hole and matter field. Here recall that section \ref{sec-eq} and reference \cite{ref-trans.1} consider the energetics of equilibrium model in which a black body of temperature $T_g$ is put in a heat bath of the same temperature $T_g$ and gravitational interaction between them is ignored. Hence section \ref{sec-eq} and reference \cite{ref-trans.1} can retain the essence of black hole phase transition. It is the self-gravitational effect of black hole on its own thermodynamic state that causes the black hole phase transition.

So far we have considered an equilibrium between a black hole and a matter field. Finally turn our discussion to the case that the matter field is not in an equilibrium with the black hole. In the nonequilibrium case it is not generally clarified whether equation \eqref{eq-doppler.1} holds or not, since nonequilibrium statistical mechanics has not yet been established in a general form. However when we consider non-self-interacting massless matter fields (named as the radiation fields in the NPT model), then we can treat the nonequilibrium effect of matter fields by two-temperature steady state thermodynamics for the radiation fields \cite{ref-sst}, which is summarised in appendix \ref{app-sst}. Hence it is expected that the original model considered in reference \cite{ref-trans.1} can be extended to include the nonequilibrium effects of Hawking radiation with utilising the steady state thermodynamics for the radiation fields.

\section{Steady state thermodynamics for the radiation fields}
\label{app-sst}

This appendix summarises two-temperature steady state thermodynamics for a radiation field \cite{ref-sst}. As done in reference \cite{ref-sst}, we consider a photon gas as the radiation field. However it is easily extended to general non-self-interacting massless matter fields by replacing the Stefan-Boltzmann constant $\sigma$ with the generalised one $\sigmap$ as mentioned at the end of section \ref{sec-intro}.

\begin{figure}[t]
 \begin{center}
 \includegraphics[height=20mm]{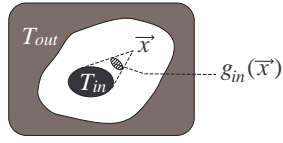}
 \end{center}
\caption{SST model. Radiation field between outer and inner black bodies is in a steady state. }
\label{fig-app.1}
\end{figure}

We consider the following model which is named SST after the Steady State Thermodynamics.
\begin{description}
\item[SST model:] Make a vacuum cavity in a large black body of equilibrium temperature $T_{out}$ and put an another smaller black body of temperature $T_{in} \, ( \neq T_{out} )$ in the cavity (see figure \ref{fig-app.1}). For the case $T_{in} > T_{out}$, the radiation field (photon gas) emitted by two black bodies causes an energy flow from the inner black body to the outer one. When the outer black body is isolated from outside environments, the whole system which consists of two black bodies and radiation field between them relaxes to a total equilibrium state in which two black bodies and radiation field have the same equilibrium temperature. 
\end{description}
Note that equations of states of two black bodies are not specified, and we assume that gravitational interactions among two bodies and radiation field is negligible. Further we consider the case satisfying the following two conditions: the first is that thermodynamic state of each black body changes along a sequence of equilibrium states during the relaxation process of the whole system, and the second condition is that the volume of cavity is so small that the speed of light is approximated as infinity during the relaxation process. Then, as shown in reference \cite{ref-sst}, thermodynamic state of radiation field sandwiched by two black bodies should change along a sequence of nonequilibrium states during the relaxation process. Further due to the Stefan-Boltzmann law, each nonequilibrium state in the sequence of radiation field's states has an energy flow $J_{sst} = \sigma \, (\, T_{in}^{\,4} - T_{out}^{\,4} \, ) \, A_{in}$, where $A_{in}$ is the surface area of inner black body. This $J_{sst}$ equals the net energy exchanged par a unit time between the two black bodies via the radiation field.

From the above discussions, it is enough for us to consider a set (state space) of steady states (macroscopically stationary nonequilibrium state) of the radiation field whose nonequilibrium nature is raised by the steady energy flow $J_{sst}$. That is, the sequence of nonequilibrium states of the radiation field during the relaxation process lies in the set (state space) of steady states. A consistent thermodynamic framework for the steady states of radiation field has already been constructed \cite{ref-sst}. The outline of the construction of internal energy and entropy are as follows. 

For the first we consider a general bosonic gas. The energy $E_b$ of the gas is given by,
\sikib
 E_b = \int \frac{dp^3}{(2\, \pi)^3} dx^3 \,
  g_{\vec{p},\vec{x}} \,\, \epsilon_{\vec{p},\vec{x}} \,\, f_{\vec{p},\vec{x}} \, ,
\label{eq-sst.ll.energy}
\sikie
where $\vec{p}$ is a momentum of particle, $\vec{x}$ is a spatial point, $\epsilon_{\vec{p},\vec{x}}$ is an energy of particle of $\vec{p}$ at $\vec{x}$, and $g_{\vec{p},\vec{x}}$ and $f_{\vec{p},\vec{x}}$ are respectively the number of states and the average number of particles at a point $(\vec{p},\vec{x})$ in the phase space of the bosonic gas. Further, refer to the Landau-Lifshitz type definition of nonequilibrium entropy $S_b$ for bosonic gas \cite{ref-ll},
\sikib
 S_b =
  \int \frac{dp^3}{(2\, \pi)^3} dx^3 \, g_{\vec{p},\vec{x}}
  \left[\, \left(\, 1 + f_{\vec{p},\vec{x}} \,\right) \,
             \ln\left(\, 1 + f_{\vec{p},\vec{x}} \,\right)
         - f_{\vec{p},\vec{x}} \, \ln f_{\vec{p},\vec{x}}
  \,\right] \, .
\label{eq-sst.ll.entropy}
\sikie
Note that it has also been shown in reference \cite{ref-ll} that the maximisation of $S_b$ for an isolated system ($\delta S_b = 0$) gives the equilibrium Bose distribution. This is frequently referred in many works on nonequilibrium systems as the {\it H-theorem}. However in reference \cite{ref-ll}, concrete forms of $g_{\vec{p},\vec{x}}$ and $f_{\vec{p},\vec{x}}$ are not specified, since an arbitrary system is considered.

In the SST model, the radiation field is sandwiched by two black bodies, and the particle (photon) is massless and {\it collisionless}. Then we can determine $\epsilon_{\vec{p},\vec{x}}$, $g_{\vec{p},\vec{x}}$ and $f_{\vec{p},\vec{x}}$ as
\sikib
 \epsilon_{\vec{p},\vec{x}} = \omega \quad , \quad
 g_{\vec{p},\vec{x}} = 2 \,\,\text{(helicities of a photon gas)} \quad , \quad
 f_{\vec{p},\vec{x}} = \frac{1}{\exp[\omega / T(\vec{p},\vec{x})] - 1} \, ,
\label{eq-sst.distribution}
\sikie
where a frequency of photon $\omega = \left| \vec{p} \right|$, and $T(\vec{p},\vec{x})$ is given by
\sikib
 T(\vec{p},\vec{x}) =
 \begin{cases}
  T_{in}  & \text{for} \,\,\, \vec{p} = \vec{p}_{in} \,\,\, \text{at $\vec{x}$} \\
  T_{out} & \text{for} \,\,\, \vec{p} = \vec{p}_{out} \,\,\, \text{at $\vec{x}$}
 \end{cases} \, ,
\sikie
where $\vec{p}_{in}$ is a momentum of photon emitted by the inner black body and $\vec{p}_{out}$ by the outer black body. The $\vec{x}$-dependence of $T(\vec{p},\vec{x})$ arises, because the directions in which photons of $\vec{p}_{in}$ and $\vec{p}_{out}$ can come to a point $\vec{x}$ vary from point to point. 

From the above, we obtain the steady state energy $E_b$ and entropy $S_b$,
\sikib
 E_b = \int dx^3 e_{rad}(\vec{x}) \quad , \quad
 e_b(\vec{x}) = 4 \, \sigma \,
  \left(\, \gin{x}\, T_{in}^{\,4} + \gout{x}\, T_{out}^{\,4} \,\right) \, ,
\label{eq-sst.energy.rad}
\sikie
and
\sikib
 S_b = \int dx^3 s_{rad}(\vec{x}) \quad , \quad
 s_b(\vec{x}) = \frac{16 \, \sigma}{3} \,
  \left(\, \gin{x}\, T_{in}^{\,3} + \gout{x}\, T_{out}^{\,3} \,\right) \, ,
\label{eq-sst.entropy.rad}
\sikie
where $\gin{x}$ is a solid angle (divided by $4 \pi$) covered by directions of $\vec{p}_{in}$ at $\vec{x}$ as shown in figure \ref{fig-app.1}, and $\gout{x}$ is similarly defined with $\vec{p}_{out}$. By definition, we have $\gin{x} + \gout{x} = 1$. Here it should be noted that, if the radiation field is in an equilibrium of temperature $T$ and volume $V$, the energy $E = 4 \sigma \, T^4 \, V$ and the entropy $S = ( 16 \sigma/3 ) \, T^3 \, V$. Hence we find that, because photons are collisionless (non-self-interacting matter field), the steady state energy and entropy are given by a simple linear combination of the values which are calculated if the radiation field is in an equilibrium of temperature $T_{in}$ or $T_{out}$. 

Further, with defining the other variables like free energy with somewhat careful discussions, it has already been checked that the 0th, 1st, 2nd and 3rd laws of ordinary equilibrium thermodynamics are extended to include the steady states of a radiation field. That is, the steady state thermodynamics for a radiation field has already been constructed in a consistent way. Especially on the steady state entropy \eqref{eq-sst.entropy.rad}, it has already been revealed in \cite{ref-sst} that the total entropy of the whole system which consists of two black bodies and radiation filed increases monotonously during the relaxation process of the whole system, $dS_{in} + dS_{out} + dS_b \ge 0$, where the equality holds for total equilibrium states, $S_{in}$ and $S_{out}$ are entropies of two black bodies which are given by ordinary equilibrium thermodynamics, and it is assumed that the two black bodies are made of ordinary materials and have positive heat capacities.

When we consider a massless fermionic field instead of the radiation field in SST model, the formula of energy \eqref{eq-sst.ll.energy} holds for fermions as well. But the formula of entropy \eqref{eq-sst.ll.entropy} is replaced by \cite{ref-ll}
\sikib
 S_f =
 -\int \frac{dp^3}{(2\, \pi)^3} dx^3 \, g_{\vec{p},\vec{x}}
  \left[\, \left(\, 1 - f_{\vec{p},\vec{x}} \,\right) \,
             \ln\left(\, 1 - f_{\vec{p},\vec{x}} \,\right)
         + f_{\vec{p},\vec{x}} \, \ln f_{\vec{p},\vec{x}}
  \,\right] \, .
\label{eq-sst.ll.fermion}
\sikie
Then, following the same procedure as for a radiation field with using the distribution function of massless fermions $f_{\vec{p},\vec{x}} = [\,\exp[\omega / T(\vec{p},\vec{x})] + 1\,]^{-1}$, we can obtain the steady state energy and entropy of a massless fermionic field as
\sikib
 E_f = \frac{7}{8} \, \frac{n_f}{2}\, E_b \quad , \quad
 S_f = \frac{7}{8} \, \frac{n_f}{2}\, S_b \, ,
\label{eq-sst.fermion}
\sikie
where $g_{\vec{p},\vec{x}} = n_f$ is the effective number of helicities of the fermions. Hence, when the black bodies in SST model emit $n_f$ massless fermionic modes and $n_b$ massless bosonic modes ($n_b = 2$ for a radiation field), the steady state energy and entropy of matter fields between two black bodies is given by equations \eqref{eq-sst.energy.rad} and \eqref{eq-sst.entropy.rad} with replacing $\sigma$ by $\sigmap$ given at the end of section \ref{sec-intro}.

Finally for the convenience to follow equations \eqref{eq-sst.energy.rad}, \eqref{eq-sst.entropy.rad} and \eqref{eq-sst.fermion}, we list useful formulae, 
\begin{flalign*}
 \int^{\infty}_0 dx \, \frac{x^3}{e^x - 1} &= \frac{\pi^4}{15} \,,
  & \int^{\infty}_0 dx \, \frac{x^2}{e^x - 1} \, \ln( e^x - 1 ) &= \frac{11 \pi^4}{180} \,,
  & \int^{\infty}_0 dx \, \frac{x^2}{1 - e^{-x}} \, \ln( 1 - e^{-x} ) &= - \frac{\pi^4}{36} \,, \\
 \int^{\infty}_0 dx \, \frac{x^3}{e^x + 1} &= \frac{7 \pi^4}{120} \,,
  & \int^{\infty}_0 dx \, \frac{x^2}{e^x + 1} \, \ln( e^x + 1 ) &= \frac{\pi^4}{60} + F \,,
  & \int^{\infty}_0 dx \, \frac{x^2}{1 + e^{-x}} \, \ln( 1 + e^{-x} ) &= \frac{11 \pi^4}{180} - F \,,
\end{flalign*}
where $F = [ (\ln 2)^2 - \pi^2 ]\,(\ln 2)^2/6 + 4 \, \phi(4,1/2) + (7 \ln 2/2)\, \zeta(3)$, $\zeta(z)$ is the zeta function and $\phi(z,s)$ is the modified zeta function (Apell's function).

\section{Heat capacity $C_X(R_g)$ as a function of $R_g$}
\label{app-CX.Rg}

\begin{figure}[t]
 \centerline{\includegraphics[height=35mm]{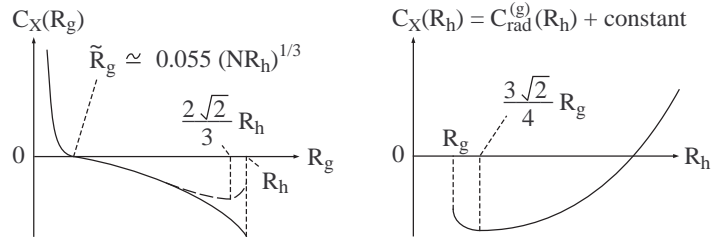}}
\caption{Left panel is for $C_X(R_g)$ as a function of $R_g$, where $2 \sqrt{2}/3 \simeq 0.943$. Right is for $C_X(R_h)$ as a function of $R_h$, where $3 \sqrt{2}/4 \simeq 1.06$. Due to the validity of NPT model, $\sigmap = O(1)$ and $O(1) < R_g \le R_h$, then $C_X(R_g = R_h) < 0$ holds.}
\label{fig-app.2}
\end{figure}

This appendix shows a behaviour of heat capacity $C_X$ of sub-system X as a function of black hole radius $R_g$. The $C_X(R_g)$ is given by equations \eqref{eq-eq.capacity} and \eqref{eq-ne.capacity} as
\sikib
 C_X(R_g) = C_g(R_g) + C_{rad}^{(g)}(R_g) \, ,
\sikie
where $0 < R_g < R_h$ by definition and
\sikib
 \begin{cases}
  C_g(R_g) = - 2 \pi\, R_g^2 \\
  C_{rad}^{(g)} = 16\,\sigmap\,G_g\,T_g^3
                + \dfrac{\sigmap}{2\,\pi}\,\left(R_g - \sqrt{R_h^2 - R_g^2} \right)\,T_g 
 \end{cases} \, .
\sikie
By equations \eqref{eq-eq.eos} and \eqref{eq-ne.G}, this is rewritten into the following form,
\sikib
 C_X(x) = - 2 \pi R_h^2\, x^2 + \frac{\sigmap}{8 \pi^2}\, f(x) \quad , \quad
 x = \frac{R_g}{R_h} \, ,
\label{eq-CX.CX}
\sikie
where
\sikib
 f(x) = \frac{4}{3}\,\frac{1}{x^3}\, \left[\, 1 - x^3 - \left( 1 - x^2 \right)^{3/2} \,\right]
      + \frac{1}{x}\,\left(\, x - \sqrt{1 - x^2} \,\right) \, .
\label{eq-CX.f}
\sikie
Then we find
\sikib
 \begin{cases}
  C_X \, \to \, \infty                           & \text{as $x \,\to\,0$} \\
  C_X = - 2 \pi R_h^2 + \dfrac{\sigmap}{8 \pi^2} & \text{at $x = 1$}
 \end{cases} \, .
\label{eq-CX.limit}
\sikie
Here recall the facts; $R_h > R_g$ by definition, $( R_h >) R_g > O(1)$ due to the quasi-equilibrium assumption, and $N = O(10)$ by equation \eqref{eq-ne.N}. Then we can expect that an inequality $C_X(x=1) < 0$ holds for the NPT model. Further the differential of $C_X(x)$ is given by
\sikib
 \frac{d C_X(x)}{dx} = - 4 \pi R_h^2\, x + \frac{\sigmap}{8 \pi^2}\, \frac{d f(x)}{dx} \, ,
\sikie
where
\sikib
 \dfrac{d f(x)}{dx} = \dfrac{4 - 3\, x^2 - 4\,\sqrt{1-x^2}}{x^4 \,\sqrt{1-x^2}}
 \quad \Rightarrow \quad
 \begin{cases}
  \dfrac{d f(x)}{dx} < 0 & \text{for $x < \dfrac{2 \sqrt{2}}{3}$} \\
  \dfrac{d f(x)}{dx} > 0 & \text{for $x > \dfrac{2 \sqrt{2}}{3}$}
 \end{cases} \, ,
\label{eq-CX.df}
\sikie
where $2\sqrt{2}/3 \simeq 0.943$. Adjusting the values of $R_h$ and $N$ in $\sigmap$, an inequality $dC_X/dx > 0$ may hold for $x > 2\sqrt{2}/3$. However, because of $C_X(x=1) < 0$, it is obvious that the equation $C_X = 0$ has only one solution. Hereafter $\tilde{R}_g\, (= R_h\, \tilde{x}$) denotes the solution of $C_X(\tilde{R}_g) = 0$. A schematic graph of $C_X(R_g)$ is shown in figure \ref{fig-app.2} (left panel).

Finally in this appendix, we estimate the value of $\tilde{x}$. Since $x < 1$ by definition, we apply the Tayler expansion, $(1-x^2)^{\alpha} = 1 - \alpha\, x^2 + O(x^4)$, to equation \eqref{eq-CX.CX} and obtain
\sikib
 C_X(x) = - 2 \pi R_h^2 \,
           \left[\, x^2 - \varepsilon \, \frac{1}{x}\,
                          \left( 1 - \frac{x}{3} + O(x^2) \right)  \,\right] \, ,
\sikie
where $\varepsilon = \sigmap/16 \pi^3 R_h^2 = N/1920 \pi R_h^2$. By equation \eqref{eq-ne.N} and $R_h > R_g > O(1)$ due to the quasi-equilibrium assumption, we find $\varepsilon < 1$. Then $C_X(x)$ becomes
\sikib
 C_X(x) = - 2 \pi R_h^2 \,\frac{1}{x}\,
          \left(\, x^3 - \varepsilon + \varepsilon\, O(x) \,\right) \, .
\label{eq-CX.0}
\sikie
Hence, by taking leading terms of $x$ and $\varepsilon$, we find an approximate expression for $\tilde{x}$ as
\sikib
 C_X(\tilde{x}) = 0 \quad \Rightarrow \quad
 \tilde{x} \simeq \varepsilon^{1/3} \, .
\sikie
This gives an approximate value of $\tilde{R}_g$,
\sikib
 \tilde{R}_g \simeq \left( \frac{N R_h}{1920 \pi} \right)^{1/3}
             \simeq 0.055 \times \left( N R_h \right)^{1/3} \, .
\label{eq-CX.tRg}
\sikie

\section{Heat capacity $C_X(R_h)$ as a function of $R_h$}
\label{app-CX.Rh}

This appendix shows a behaviour of heat capacity $C_X$ of sub-system X as a function of outermost radius $R_h$ of the hollow region in NPT model. We make use of the calculations done in appendix \ref{app-CX.Rg}. According to equation \eqref{eq-CX.CX}, $C_X(R_h)$ is expressed as
\sikib
 C_X(R_g) = - 2 \pi R_g^2 + C_{rad}^{(g)}(R_h) \quad , \quad
 C_{rad}^{(g)}(R_h) = \frac{\sigmap}{8 \pi^2} \, f(x) \, ,
\sikie
where $x = R_g/R_h$, $0 < x < 1$ by definition, and $f(x)$ is given by equation \eqref{eq-CX.f}. This denotes that $C_X(R_h)$ behaves as $C_{rad}^{(g)}(R_h) + constant$. The differential becomes
\sikib
 \frac{d C_X(R_h)}{dR_h} = - \frac{\sigmap}{8 \pi} \, x^2 \, \frac{d f(x)}{dx} \, .
\sikie
Using equation \eqref{eq-CX.df}, we find
\sikib
 \begin{cases}
  \dfrac{dC_X}{dR_h} = \dfrac{\sigmap}{8 \pi^2} & \text{at $R_h = R_g$ ($x = 1$)} \\[3mm]
  \dfrac{dC_X}{dR_h} \to - \infty               & \text{as $R_h \to \infty$ ($x \to 0$)}
 \end{cases} \, .
\sikie
Hence referring to the limit values \eqref{eq-CX.limit} and the behaviour \eqref{eq-CX.df} of $df/dx$, a schematic graph of $C_X(R_h)$ is obtained as shown in figure \ref{fig-app.2} (right panel). This denotes that, since $C_X < 0$ must hold by the validity of NPT model \eqref{eq-ne.validity.1}, the absolute value $\left| C_X(R_h) \right|$ is monotone increasing for $R_h > (3 \sqrt{2}/4) R_g \simeq 1.06 R_g$ in the framework of NPT model.

\section{Proof of the inequality $C_g/C_X > 1$}
\label{app-Cg/CX}

This appendix gives a proof of inequality $C_g/C_X > 1$ under the condition $C_X < 0$ (see the discussion on inequality \eqref{eq-ne.validity.1} ). We make use of the calculations done in appendix \ref{app-CX.Rg}. By equation \eqref{eq-CX.CX} and definition of $C_X$ given in equations \eqref{eq-ne.capacity}, the heat capacity $C_{rad}^{(g)}$ of out-going radiation fields is expressed as
\sikib
 C_{rad}^{(g)}(x) = \frac{\sigmap}{8 \pi^2} \, f(x) \, ,
\sikie
where $x = R_g/R_h$, $0 < x < 1$ by definition, and $f(x)$ is given by equation \eqref{eq-CX.f}. Then equation \eqref{eq-CX.df} indicates $C_{rad}^{(g)}(x) \ge C_{rad}^{(g)}(2\sqrt{2}/3)$. Here equation \eqref{eq-CX.f} gives $f(2\sqrt{2}/3) \simeq 0.845 \, \Rightarrow \, f(2\sqrt{2}/3) > 0$. Therefore we find $C_{rad}^{(g)} > 0$ for $0 < x <1$, which is consistent with a naive expectation that an ordinary matter like radiation fields has generally a positive heat capacity. Consequently the condition $C_X < 0$ gives an inequality $C_g/C_X > 1$ for $0 < x < 1$ as follows (see equation \eqref{eq-eq.capacity} for $C_g$),
\sikib
 C_X = C_g + C_{rad}^{(g)} < 0 \quad \Rightarrow \quad
 0 < C_{rad}^{(g)} < \left| C_g \right| \quad \Rightarrow \quad
 \left| C_X \right| = \left| C_g \right| - C_{rad}^{(g)}
                    < \left| C_g \right| \quad \Rightarrow \quad
 \frac{C_g}{C_X} > 1 \, .
\sikie

\section{Combined heat capacity $C_X + C_Y^{(g)}$ as a function of $R_g$}
\label{app-CXCYg}

This appendix shows a behaviour of combined heat capacity $C_X + C_Y^{(g)}$ as a function of black hole radius $R_g$. We make use of the calculations done in appendix \ref{app-CX.Rg}. By equation \eqref{eq-CX.CX} and definitions of $C_X$ and $C_Y^{(g)}$ given in equations \eqref{eq-ne.capacity}, $C_X + C_Y^{(g)}$ is expressed as
\sikib
 C_X(R_g) + C_Y^{(g)}(R_g) =
 - 2 \pi R_h^2\, x^2 + \frac{\sigmap}{8 \pi^2}\,
 \left[\, f(x) + q\, x^3 \,\left(\, x + \sqrt{1-x^2} \,\right) \,\right] \, ,
\sikie
where $q = \left( 4 \pi R_h \, T_h \right)^4$, $x = R_g/R_h$, $0 < x < 1$ by definition, and $f(x)$ is given by equation \eqref{eq-CX.f}. Then, using the limit values \eqref{eq-CX.limit}, we find
\sikib
 \begin{cases}
  C_X + C_Y^{(g)} \, \to \, \infty                           & \text{as $x \,\to\,0$} \\
  C_X + C_Y^{(g)} =
    - 2 \pi R_h^2 + \dfrac{\sigmap}{8 \pi^2} + \dfrac{\sigmap}{8 \pi^2} \, q & \text{at $x = 1$}
 \end{cases} \, .
\sikie
The first two terms in $C_X + C_Y^{(g)}$ at $x=1$ is negative in the framework of NPT model as discussed in equations \eqref{eq-CX.limit}. Therefore, if $T_h$ is sufficiently small, then $q$ can be small enough so that $C_X + C_Y^{(g)}$ at $x=1$ becomes negative.

\begin{figure}[t]
 \begin{center}
  \includegraphics[height=35mm]{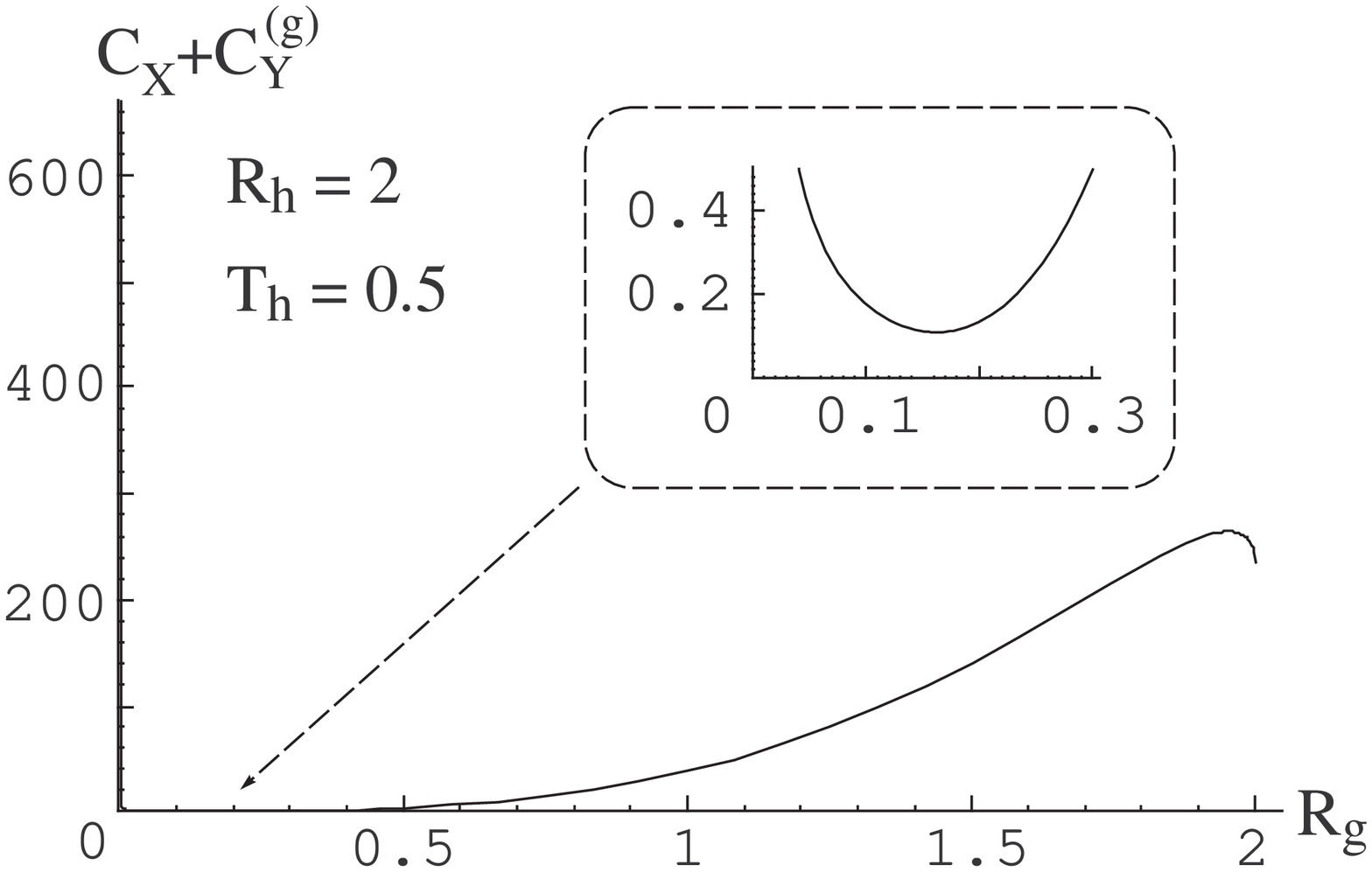}
  \includegraphics[height=35mm]{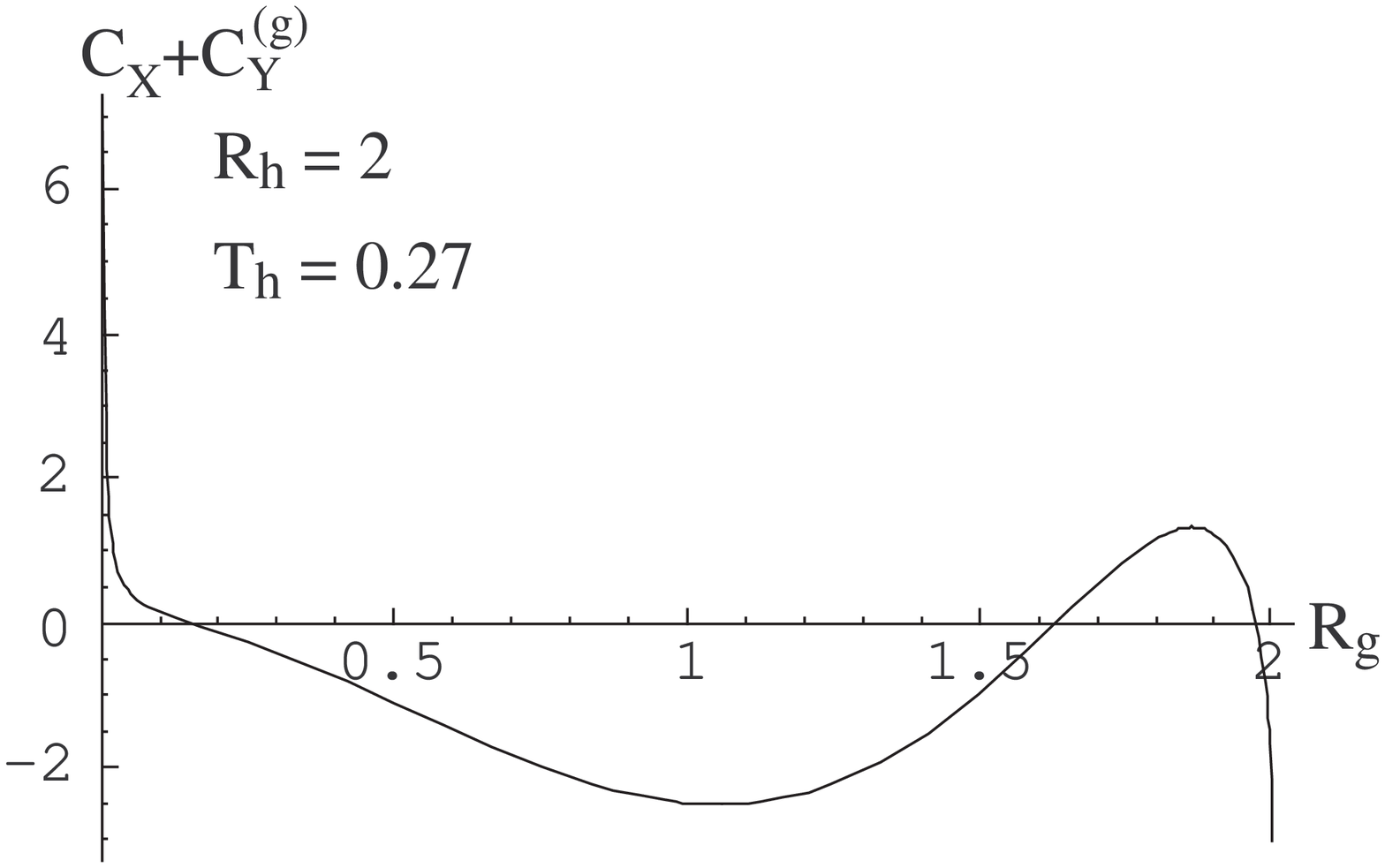}
  \includegraphics[height=35mm]{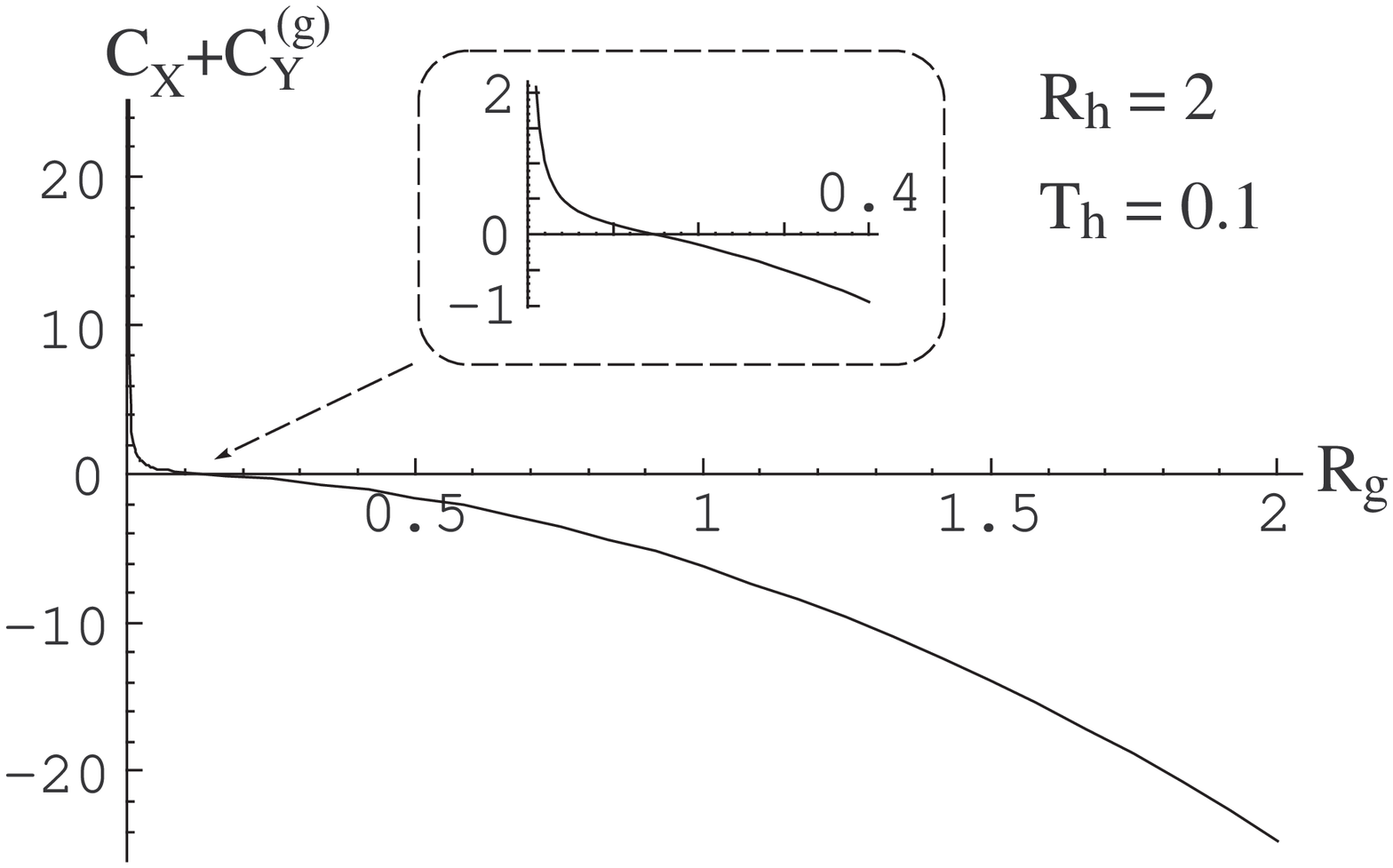} \\
  \includegraphics[height=35mm]{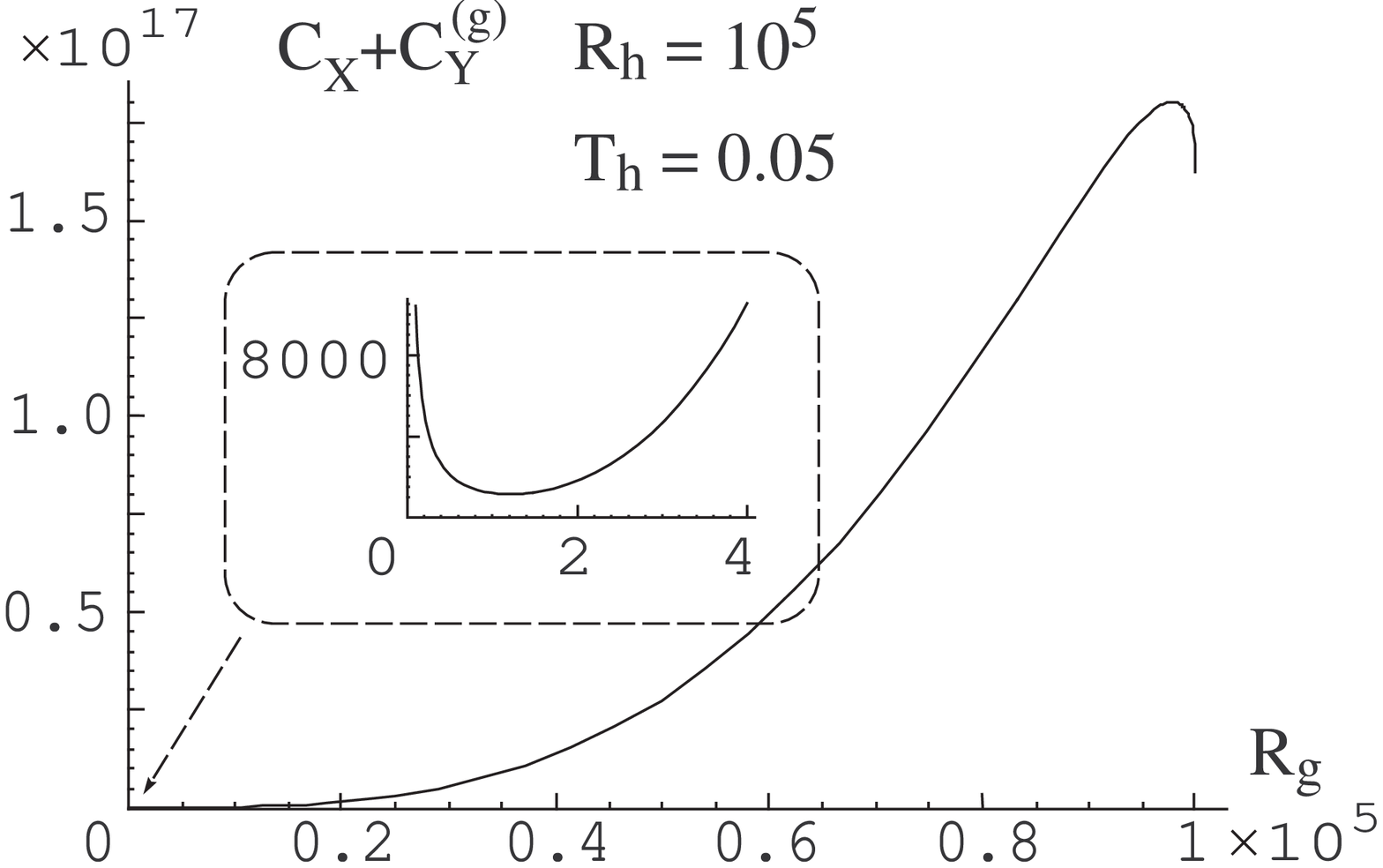}
  \includegraphics[height=35mm]{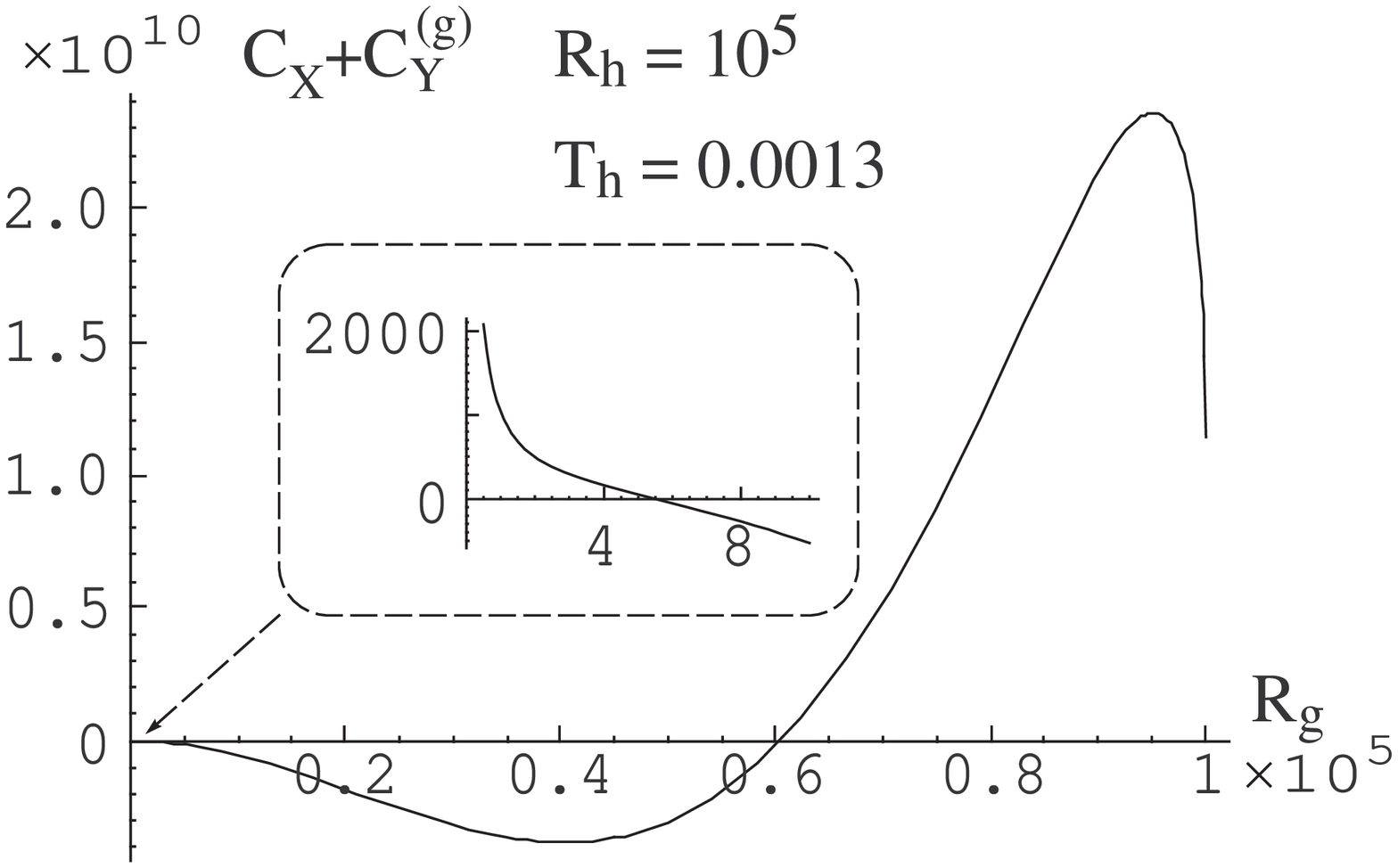}
  \includegraphics[height=35mm]{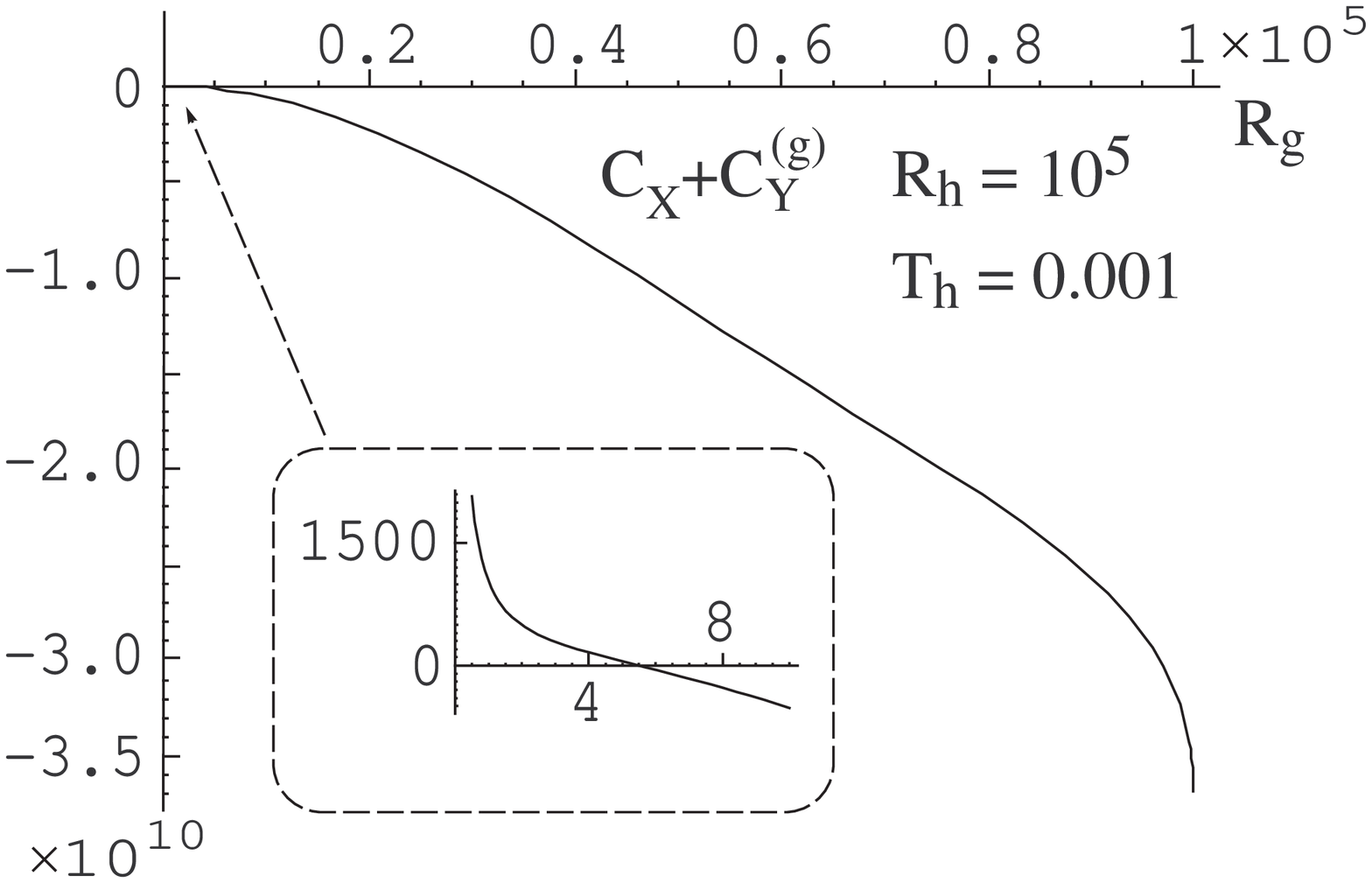}
 \end{center}
\caption{Numerical plots of $C_X + C_Y^{(g)}$ as a function of $R_g$ with $N = 10$. Upper three panels are for $R_h = 2$ and $T_h = 0.5$, $0.27$ and $0.1$ from left to right. Lower three panels are for $R_h = 10^5$ and $T_h = 0.05$, $0.0013$ and $0.001$ from left to right. It has been found for every value of $R_h > 1$ as far as the author checked that the same behaviour is observed with decreasing $T_h$, and $C_X + C_Y^{(g)} < 0$ can hold for a sufficiently small $T_h$.}
\label{fig-app.3}
\end{figure}

The differential of $C_X + C_Y^{(g)}$ is complicated and not suitable for analytical utility. Instead of analytical discussion, we show some numerical examples of $C_X + C_Y^{(g)}$ in figure \ref{fig-app.3}. It is recognised with this figure that, for a sufficiently small $T_h$, the condition \eqref{eq-ne.validity.1}, $C_X + C_Y^{(g)} < 0$, which guarantees the validity of NPT model can be satisfied for a certain range of $R_g$. Although figure \ref{fig-app.3} shows only some examples, it has been found as far as the author checked that the same behaviour is observed for every value of $R_h > 1$. That is, $C_X + C_Y^{(g)} < 0$ can hold for a sufficiently small $T_h$.  Here note that, during a semi-classical and quasi-equilibrium stage of black hole evaporation, it is reasonable to require $T_h < 1$ (= Planck temperature ). Hence we assume throughout the paper that $T_h$ is small enough so that the equation $C_X(R_g) + C_Y^{(g)}(R_g) = 0$ has one, two or three solutions of $R_g$.

Finally in this appendix, we discuss what happens along a black hole evaporation process for the case that the equation $C_X(R_g) + C_Y^{(g)}(R_g) = 0$ has two or three solutions of $R_g$. For the first consider the case of three solutions, and denotes these solutions by $R_{g0}^{(s)}$, $R_{g0}^{(m)}$ and $R_{g0}^{(l)}$ in an increasing order ($R_{g0}^{(s)} < R_{g0}^{(m)} < R_{g0}^{(l)}$). When the evaporation starts with initial black hole radius $R_g(0)$ which is in the range $R_{g0}^{(l)} < R_g(0) < R_h$ (see for example upper-centre panel in figure \ref{fig-app.3}), then the evaporation process can be treated in the framework of NPT model until the radius $R_g(t)$ decreases to $R_{g0}^{(l)}$. According to the discussion in section \ref{sec-ne.energy} (the paragraph one before the last one), the evaporation process can not be treated in the framework of NPT model after the black hole radius decreases to $R_{g0}^{(l)}$, because the validity condition \eqref{eq-ne.validity.1} of NPT model is violated in the range $R_{g}^{(m)} < R_g < R_{g0}^{(l)}$. However it is reasonable to expect that, even if the NPT model is not applicable to analyse the evaporation process while the black hole radius decreases from $R_{g}^{(l)}$ to $R_{g0}^{(m)}$, the NPT model becomes valid again after the black hole radius reaches $R_{g0}^{(m)}$. That is, the evaporation process can be analysed in the framework of NPT model in the range $R_{g0}^{(s)} < R_g < R_{g0}^{(m)}$.

Next consider the case that the equation $C_X(R_g) + C_Y^{(g)}(R_g) = 0$ has two solutions. When the black hole evaporation starts with initial radius which is larger than the larger solution of $C_X(R_g) + C_Y^{(g)}(R_g) = 0$ (see for example lower-centre panel in figure \ref{fig-app.3}), then the similar discussion as given in previous paragraph can be applied. That is, although the black hole evaporation process can not be treated in the framework of NPT model until the black hole radius becomes smaller than the larger solution of $C_X(R_g) + C_Y^{(g)}(R_g) = 0$, it becomes possible that the evaporation process can be analysed by the NPT model until the radius decreases to the smaller solution of $C_X(R_g) + C_Y^{(g)}(R_g) = 0$.

From the above we find that, for both of the cases that the equation $C_X(R_g) + C_Y^{(g)}(R_g) = 0$ has two and three solutions, it is the lowest solution of the equation $C_X(R_g) + C_Y^{(g)}(R_g) = 0$ at which the evaporation process comes not to be treated in the framework of NPT model and the evaporation process proceeds to a highly nonequilibrium dynamical stage. (According to section \ref{sec-ne.evapo}, the highly nonequilibrium dynamical stage is the abrupt catastrophic evaporation at semi-classical level for the case $R_h \gtrsim O(10^3)$.)

\section{On the final stage of evaporation: except for the NPT model}
\label{app-final}

This appendix shows that the abrupt catastrophic evaporation at semi-classical level does not occur for the equilibrium model used in section \ref{sec-eq} and for the model that a black hole evaporates in an empty space. 

For the first consider the equilibrium model used in section \ref{sec-eq}. Since there is no nonequilibrium region between black hole and heat bath, the energy transport equations are given by
\sikib
 \frac{d E_g}{dt} = - \sigmap \left(\, T_g^4 - T_h^4 \,\right) A_g \quad , \quad
 \frac{d E_h}{dt} = \sigmap \left(\, T_g^4 - T_h^4 \,\right) A_g \, ,
\sikie
where $A_g = 4 \pi R_g^2$ is the area of black hole surface. The first equation of $dE_g/dt$ is used in the following analysis. Then one of equations of states \eqref{eq-eq.eos}, $E_g = R_g/2$, gives
\sikib
 v \equiv \left| \frac{d R_g}{dt} \right| = 2 \left| \frac{d E_g}{dt} \right|
   = 2 \sigmap \left(\, T_g^4 - T_h^4 \,\right) A_g \, .
\sikie
Here note that, since the equilibrium model is based on equations of states \eqref{eq-eq.eos}, the quasi-equilibrium assumption is also necessary for this model and $v < 1$ is required. It is obvious that $v = 1$ occurs for non-zero radius $R_g > 0$, since $T_g^4 \, A_g \propto 1/R_g^2 \to \infty$ as $R_g \to 0$. The inequality $v < 1$ corresponds to the validity condition \eqref{eq-ne.validity.2} of NPT model. On the other hand, for the heat capacity $C_g$ of black hole, it is obvious that $C_g = 0 \Rightarrow R_g = 0$. That is, there is no counterpart in the equilibrium model which corresponds to the validity condition \eqref{eq-ne.validity.1} of NPT model. Hence we can recognise that, if $v = 1$ occurs for a black hole of semi-classical size $R_g > 1$, it is reasonable to conclude that the abrupt catastrophic evaporation at semi-classical level occurs. And if $v = 1$ does not occur for a semi-classical black hole, then the abrupt catastrophic evaporation does not occur in the framework of equilibrium model. 

The validity condition $v < 1$ is rewritten as follows,
\sikib
 \sigmap \left(\, T_g^4 - T_h^4 \,\right) < 2 \pi \, T_g^2 \quad \Rightarrow \quad
 T_g^4 - \beta\, T_g^2 - T_h^4 < 0 \quad \Rightarrow \quad
 R_g^2 > \frac{1}{8 \pi^2 \beta}
         \left(\, 1 + \sqrt{1 + \left( 2 T_h^2/\beta \right)^2} \,\right)^{-1} \, ,
\label{eq-final.eq}
\sikie
where $\beta = 2 \pi/\sigmap = 240/\pi N$ and $R_g = 1/4 \pi T_g$ is used. Here equation \eqref{eq-ne.N} gives $\beta = O(10)$. Further note that $T_g > T_h$ holds generally for any evaporation process and that $T_g < 1$ holds due to the quasi-equilibrium assumption. Then we find $T_h < 1$ and $(2 T_h^2/\beta)^2 \ll 1$. Hence we can approximate inequality \eqref{eq-final.eq} to the following form
\sikib
 R_g^2 > \frac{1}{8 \pi^2 \beta} \quad \Rightarrow \quad R_g \gtrsim O(0.01) \, ,
\sikie
where $\beta = O(10)$ is used. This denotes that the fast evaporation $v = 1$ occurs at $R_g \sim 0.01$. Hence, as discussed in previous paragraph, it is reasonable to conclude that the abrupt catastrophic evaporation does not occur in the framework of equilibrium model used in section \ref{sec-eq}. 

Next consider a black hole evaporation in an empty space. Ignoring gravitational interactions between black hole and radiation fields (curvature scattering, lens effects and so on), the Stefan-Boltzmann law gives (see the energy emission rate $\je$ defined by equation \eqref{eq-ne.rate.empty})
\sikib
 \frac{d E_g}{dt} = - \sigmap T_g^4 A_g \, .
\sikie
Then one of equations of states \eqref{eq-eq.eos}, $E_g = R_g/2$, gives
\sikib
 v = \left| \frac{d R_g}{dt} \right| = 2 \left| \frac{d E_g}{dt} \right|
   = 2 \sigmap T_g^4 A_g \, .
\sikie
Hence, following a similar calculation given in previous paragraph with setting $T_h = 0$, we obtain
\sikib
 v < 1 \quad \Rightarrow \quad
 R_g^2 > \frac{1}{8 \pi^2 \beta} = O(10^{-3}) \quad \Rightarrow \quad
 R_g \gtrsim O(0.01) \, .
\sikie
This denotes that the fast evaporation $v = 1$ occurs at $R_g \sim 0.01$, and that it is reasonable to conclude that the abrupt catastrophic evaporation does not occur for a black hole evaporation in an empty space with ignoring gravitational interactions between black hole and radiation fields.



\end{document}